\documentclass[11pt,titlepage]{article}
\usepackage{amsmath,amsfonts,amssymb,color}
\usepackage{graphicx}
\usepackage[round,numbers,sort&compress]{natbib} 
\usepackage{fullpage}

\newcommand{\Red}[1]{{\color{black}#1}}
\newcommand{\Blue}[1]{{\color{black}#1}}

\title{Exploring the energy landscapes of protein folding simulations with Bayesian computation}

\author{Nikolas S. Burkoff, Csilla V\'{a}rnai, Stephen A. Wells and David L. Wild \footnote{Corresponding Author. Address: Systems Biology Centre, University of Warwick, Coventry, UK, D.L.Wild@warwick.ac.uk}}  

\date{}

\begin{document}

\maketitle

\abstract{Nested sampling is a Bayesian sampling technique developed to explore probability distributions localised in an exponentially small area of the parameter space. The algorithm provides both posterior samples and an estimate of the evidence (marginal likelihood) of the model. The nested sampling algorithm also provides an efficient way to calculate free energies and the expectation value of thermodynamic observables at any temperature, through a simple post-processing of the output. Previous applications of the algorithm have yielded large efficiency gains over other sampling techniques, including parallel tempering. 
  In this paper we describe a parallel implementation of the nested sampling algorithm and its application to the problem of protein folding in a G\={o}-\Red{like} force field of empirical potentials that were designed to stabilize secondary structure elements in room-temperature simulations. We demonstrate the method by conducting folding simulations on a number of small proteins which are commonly used for testing protein folding procedures. A topological analysis of the posterior samples is performed to produce energy landscape charts, which give a high level description of the potential energy surface for the protein folding simulations. These charts provide qualitative insights into both the folding process and the nature of the model and force field used.}


\clearpage

\section{Introduction}
Approximately 50 years ago, Anfinsen and colleagues demonstrated that
protein molecules can fold into their three-dimensional native state reversibly, leading to the view that
these structures represented the global minimum of a rugged funnel-like energy landscape \cite{Anfinsen:1973ptg, Bryngelson:1987sgs, onuchic}. 


\Red{According to the hierarchical folding theory of Baldwin and Rose, a protein folds by first forming local structural elements, $\alpha$-helices and $\beta$-strands. These secondary structure elements then interact with each other resulting in the formation of the folded protein \cite{Baldwin:1999cy, Baldwin:1999jy}. The formation of local structural elements reduces the entropy of the protein, for example, the side-chains of helical residues are strongly constrained by the rest of the helix. This loss of entropy is compensated by favourable short-range interactions, including hydrogen bonding and desolvation of backbone polar groups. This is considered to be a fundamental property of proteins, and any model system attempting to simulate protein folding should mimic this property.}    

\Red{Whilst there has been recent evidence of hierarchical folding in long time-scale molecular dynamics simulations made possible by the use of custom designed supercomputers \cite{Lindorff-Larsen}, simplified G\={o}-type models remain an important class of protein models in the investigation of energy landscapes. G\={o} models assume that non-native interactions do not contribute to the overall shape of the folding energy surface \cite{Go:1983bk,takada}. In this work we use an extended G\={o}-type model, in which a G\={o} potential captures interactions between contacts of the native state of the protein, but attractive non-native interactions are also permitted (for example, hydrogen bonds can form between residues that are not in contact in the native state). This addition allows us to explore a more realistic rugged energy landscape compared to the `perfect funnel' found in a standard G\={o} model \cite{takada}, whilst maintaining the ability to perform simulations with limited computational resources.} 


The energy landscapes of protein folding simulations are most commonly visualised in terms of two- or three-dimensional plots of microscopic or free energy versus a
`reaction coordinate', such as the fraction of residue contacts in common with the native state or the root mean squared deviation between a given conformation and the native state \cite{dinner2000understanding,sali1994does}. Originally developed for reduced lattice models, these approaches have since been used for all-atom off-lattice simulations, although, in these more realistic models, they offer only an indirect visualization of the energy landscape at a single scale \cite{clementi2003interplay}. Projection into the space defined by principal components analysis of the contact map has also been used to provide a two-dimensional visualization of the energy surface \cite{hori2009folding}. Techniques adapted from robotic motion planning have been used to provide a `probabilistic roadmap' of protein folding, which may be mapped onto a conceptual drawing of the potential energy surface \cite{amato2003using}. Protein potential energy surfaces and folding funnels have also been visualized by  disconnectivity graphs \cite{dis}  and scaled disconnectivity graphs \cite{pes, scale}. Although these latter methods have the advantage of providing a visualisation of the whole  energy landscape, they rely on creating a large database of local energy minima of the surface, and are thus impractical for large systems, and do not provide information about the  entropy of the system, which governs the widths of the conceptual protein folding funnel.

\Red{The funnel-like nature of the energy landscape provides a challenging conformational space for computer simulations to explore, because only an exponentially small number of conformations have low energy and low entropy and are found towards the bottom of the funnel; the system also undergoes a first order phase transition as the protein collapses into its native state.  In this work we use nested sampling} \Blue{to explore the energy landscapes of protein folding simulations.}
Nested sampling is 
a Bayesian sampling technique introduced by Skilling \cite{sivia,sk1}, designed to explore probability distributions where the posterior mass is localised in an exponentially small area of the parameter space. It both provides an estimate of the \textit{evidence} (also known as the \textit{marginal likelihood}, or \textit{partition function}) and produces samples of the posterior distribution. Nested sampling offers distinct advantages over  methods such as simulated annealing \cite{kirkpatrick1983optimization}, Wang-Landau sampling \cite{wang}, parallel tempering (replica exchange)  \cite{par} \Red{and annealed importance sampling \cite{potts}}, in systems 
characterized by first order phase transitions \cite{sivia,gab}. The technique reduces multidimensional problems to one dimension and has a single key parameter in the trade-off between cost and accuracy. \Blue{The calculation of free energies by thermodynamic integration \cite{bennett1976efficient} and thermodynamic observables, such as heat capacities,  typically involves multiple simulations at different temperatures. Nested sampling} provides an efficient framework for computing the partition function and hence thermodynamic observables at any temperature, without the need to generate new samples at each temperature. Hence, it allows us to directly investigate the macroscopic states of the protein folding pathway and evaluate the associated free energies. Nested sampling has previously been used in the field of astrophysics \cite{astro} and for exploring potential energy hypersurfaces of Lennard-Jones atomic clusters \cite{gab}, yielding large efficiency gains over parallel tempering. 
Its use in this paper, represents, to our knowledge, the first application of this technique to a biophysical problem.

\section{Materials and Methods}

\Blue{In general, the energy of a polypeptide, $E(\Omega,\theta)$, is defined by its
conformation, $\Omega$, and arbitrary interaction parameters, $\theta$. These 
interaction parameters may be as diverse as force constants, distance cut-offs,
dielectric permittivity, atomic partial charges, etc. This
energy, in turn, defines the probability of a particular conformation, $\Omega$, at inverse thermodynamic temperature $\beta$ via the Boltzmann
distribution:    
\begin{eqnarray}
P(\Omega,\theta| \beta ) &=& \frac{1}{Z(\theta ,\beta)}
\exp \left[ -E(\Omega,\theta) \beta \right] \label{eq:04} \\
Z(\theta , \beta) &=& \int {d\Omega} \exp \left[ -E(\Omega,\theta) \beta \right],  \label{eq:05}
\end{eqnarray}
where $Z(\theta,\beta)$ is the partition function (or \textit{evidence}, in Bayesian terminology). In the following,  energy is expressed in units of $RT$, the product of
the molar gas constant and absolute temperature and $\beta = 1/RT$.}

In Bayesian statistics, with $\theta$ an unknown parameter, $D$ the observed data and $H$ the underlying model or hypothesis, we have the following relation (Bayes' rule): Posterior x Evidence = Likelihood x Prior:
\[ \mathcal{P}(\theta|D,H)Z = \mathcal{P}(D|H,\theta)\mathcal{P}(\theta|H), \]
where $Z$, the evidence, is defined as
\[ Z = \int\mathcal{P}(D|H,\theta)\mathcal{P}(\theta|H)d\theta \mbox{ }.\]
Nested sampling provides an algorithm for estimating the evidence, $Z = \mathcal{P}(D|H)$,  and the procedure additionally explores the posterior distribution, allowing its properties to be estimated.

\subsection{Procedure}

\begin{figure}[!ht]
\begin{center}
\includegraphics[width=3.25in]{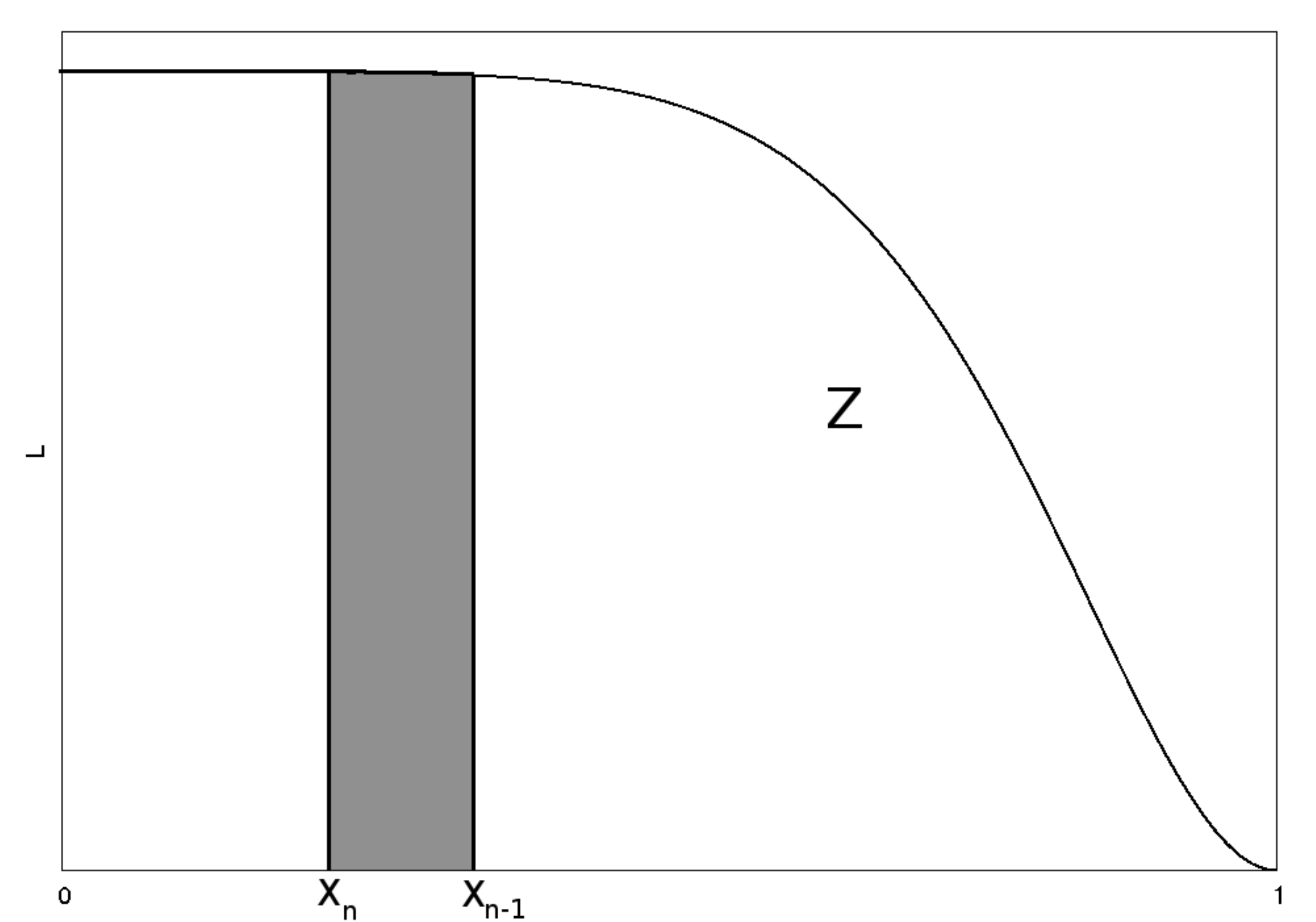}
\caption{ The evidence $Z$ is the area under the function $L(X)$. The sample $\theta_n$ represents $X_{n-1}-X_{n}$ of the  phase space volume, the proportion of the $x$-axis shaded grey. Its weighting for the posterior is $L_n(X_{n-1}-X_n)/Z$, the proportion of $Z$ shaded grey.}
\label{weight}
\end{center}
\end{figure}

We define $X(\lambda) =  \lambda$ to be the proportion of the prior with likelihood $L(X)$ greater than $\lambda$. Then, following \cite{sk1}, the evidence is:
$ Z= \int_0^1 L(X)dX$, where $L(X(\lambda)) = \lambda$ and $dX = \pi(\theta)d\theta$, with $\pi(\theta)$ the prior distribution. Fig.~\ref{weight} shows the graph of $L$ against $X$ (this is not to scale as normally the bulk of the posterior is in an exponentially small area of the phase space). $L$ is a decreasing function of $X$, as the restriction on the likelihood becomes tighter as $\lambda$ increases. The area under the curve is $Z$. The nested sampling procedure estimates points on this curve (see Algorithm below) and then uses numerical integration to calculate $Z$.

\noindent \textbf{Algorithm}
\begin{enumerate}
\item Sample (uniformly w.r.t.\ the prior) $K$ points of the parameter space $\{\theta_1 \ldots \theta_K \}$, the `active list'; calculate their likelihoods: $\{L(\theta_1),\ldots,L(\theta_K)\}$ 
\item Take the sample point with the smallest likelihood; save it as $(L_1,X_1)$
(see below for an estimate of $X$); remove this point from the active list
\item Generate a new point $\theta$ sampled uniformly (w.r.t.\ the prior) from those points
with likelihood  $L(\theta)>L^*=L_1$; add it to the active list 
\item Repeat steps 2 and 3 generating $ (L_2,X_2),(L_3,X_3),\ldots,(L_i,X_i),\ldots$ 
\end{enumerate} 

$X_1$ is located at the largest of $N$ numbers uniformly distributed on $(0,X_0)$, where $X_0 = 1$. Skilling suggests using the expected value of the shrinkage ratio,  $X_i/X_{i-1}$, to estimate $X_i$ (the estimate of $X$ for iteration $i$), where  $X_i$ is the largest of $N$ numbers uniformly distributed on $(0,X_{i-1})$ \cite{sk1}. The shrinkage ratio has probability density function
$f(t) = Kt^{K-1}$, 
with mean and standard deviation
$\log(t)= (-1\pm1)/K$, 
and, as each shrinkage ratio is independent, we find, if uncertainties are ignored,
\[ \log(X_i) = (-i\pm \sqrt{i})/K \, \Rightarrow \, X_i  \approx \exp(-i/K).\]
It is also possible to use the arithmetic expected value to estimate $X_i$ \cite{mckay}. This implies that $X_i = \alpha^i$ where $\alpha = {K}/{(K+1)}$. In the limit of large $K$, these two approaches are identical and henceforth we will use $\alpha= \exp(-1/K)$ or ${K}/{(K+1)}$, and $X_n = \alpha^n$.

\subsection{Parallel Nested Sampling}

For high dimensional systems sampling uniformly (conditional upon the likelihood being above a fixed value, $L^*$) is not computationally tractable. In this case a Markov chain can be used to explore the parameter space \cite{sivia}. To generate a new point, one of the active set of points (not necessarily the one with the lowest likelihood) is chosen to be the start of a short Monte Carlo (MC) run, with all moves that keep the likelihood above $L^*$ being accepted. 

Starting the MC run from a copy of one of the points of the active set, chosen at random, is crucial to nested sampling. Suppose we have a bimodal likelihood function. Once $L^*$ is sufficiently high, the region of the parameter space the chain is allowed to explore will no longer be connected; it will have two disconnected components. Without copying, all active points which enter the subordinate component will be trapped there. With copying, provided at least one enters the dominant mode, then as $L^*$ increases, active points in the subordinate mode will be replaced by ones from the dominant mode. This is particularly important for likelihood functions for which the dominant mode splits again at a higher likelihood. In general, for a given $K$, 
if the relative phase space volume of a mode is less than $1/K$ in comparison to the rest of the space at the splitting likelihood, the chances of nested sampling exploring the mode is small \cite{gab}. Therefore, the parameter $K$ controls the resolution of the exploration. 

The number of trial MC moves per nested sampling iteration, $m$, is another key parameter when using nested sampling for higher dimensional systems. If $m$ is too small, the parameter space is inadequately explored; new active set samples and the current conformations they are copied from remain very similar. Setting $m$ too high results in longer than necessary runtimes, as conformations part way through the MC run are already sufficiently different from their starting positions. \Red{Hence $K$ controls which regions of the parameter space are available to the algorithm and $m$ controls how well these regions are explored.}   

We parallelized the nested sampling algorithm by removing the $P$ points with the lowest likelihood  at each nested sampling iteration, one for each processor used. Each processor then runs its own independent MC simulation to replace one of the removed points. For post-processing, at each iteration we only store the point which has the $P$th lowest likelihood and adjust $\alpha$ accordingly; $\alpha=1-P/(K+1)$.

Running a parallel nested sampling algorithm with $K$ points explores the parameter space more effectively than $P$ serial nested sampling simulations each with $K/P$ points in the active set, while requiring equal computational resources.  Consider a likelihood function, which splits $n$ times in the dominant mode (i.e.\ contains the majority of the evidence), with the probabilities of an exploratory active point falling into the dominant mode being $W_1, W_2,  ... W_n$ at the critical likelihood (the likelihood of splitting).  Defining \emph{success} as exploring the dominant mode at the $n$th split in at least one simulation, it can be shown, using an argument similar to that of~\cite{sivia} that 
\begin{equation}
\mathbb{P}(\text{\small{success}} \; | \text{\small{1 simulation with $K$ points}}) = \Pi_g \left[ 1 - \left( 1 - W_g \right)^K \right]	\label{dominant-mode-prob}
\end{equation} 
and
\begin{align}
\mathbb{P}(&\text{\small{success}} \; | \text{\small{$P$ simulations with $K/P$ points}}) = \nonumber \\
&=1 - \left( 1 - \Pi_g \left( 1 - \left( 1 - W_g \right)^{K/P} \right) \right)^P . \nonumber
\end{align}

For example, if $n=2$, $W_1=W_2=0.1$, $K=32$ and $P=4$ then  $\mathbb{P}(\mathrm{success} | \mathrm{parallel} ) = 0.933$ and $\mathbb{P}(\mathrm{success} | \mathrm{serial} ) = 0.792$.


\subsection{Posterior Samples}

The sample points removed from the active set, labelled $\theta_1,\theta_2,\ldots$, say, can be used to estimate properties of the posterior distribution. Sample point $\theta_n$ represents 
\begin{equation*} \omega_n =  X_{n-1}-X_{n} \label{omega} \end{equation*} 
of the phase space volume (with respect to the prior distribution) and hence 
\begin{equation*}  \chi_n = \frac{(X_{n-1}-X_{n})L(\theta_n)}{Z} \label{chi} \end{equation*}
is the relative volume of the posterior space that $\theta_n$ represents; see Fig.~\ref{weight}. 

In the case of a Boltzmann distribution, at inverse temperature $\beta$, $L(\theta_n) = \exp(-E_n \beta)$ and hence, by calculating $\chi_n(\beta)$, a single nested sampling simulation can provide the expectation value of any thermodynamic observable, such as heat capacity, at any temperature.  Given a property $Q(\theta | \beta)$ of the posterior:
\begin{equation}
\mathbb{E}(Q | \beta ) \approx \sum_i \chi_i (\beta) Q(\theta_i).
\label{eq:thd}
\end{equation}

In energetic terms, the nested sampling scheme is built from a set of decreasing energy levels, $\{E_n\}$, with the energy of conformation $\Omega_n$ given by equation \ref{eq:12}. Each energy level has  an associated weight, which is also decreasing. At each energy level, a set of $K$ sample points (or conformations), $\{\Omega^i_n\}$, is obtained by uniform sampling from the energy landscape \emph{below} $E_n: \mbox{   }\Omega^i_n \sim U({\Omega:E(\Omega) < E_n})$.
After every iteration, a new lowest energy level $E_{n+1}$ is defined to be at a fixed fraction, $\alpha$, of the current energy distribution. In this way, a fraction $\alpha^n$ of the whole phase space has energy below $E_n$, and a fraction $\alpha^{n+1}$ has energy below $E_{n+1}$. The phase space volume will therefore shrink exponentially, by a factor of $\alpha$, with every nested sampling iteration, and the algorithm is able to locate exponentially small regions of phase space.

\subsection{The Protein Model}
\label{mod}

The polypeptide model we use is adapted from our previous published work \cite{Podtelezhnikov:2009,Podtelezhnikov:2008gm,Podtelezhnikov:2007il, Podtelezhnikov:2005so}. It is fully described in Appendix A and a summary is provided below.
  
Our polypeptide model features all-atom representations of the polypeptide
backbone and $\beta$-carbon atoms. Other side-chain atoms are represented by one or, in the case of branched side chains, two pseudo-atoms, following \cite{LINUS}.

For a given protein sequence, $R$, the Boltzmann distribution defines the probability, $P(R,\Omega |\beta)$, of it adopting a particular conformation, $\Omega$, at inverse thermodynamic temperature $\beta$. This probability can be factorized
into the product of the sequence-dependent  likelihood for a given conformation and the
prior distribution of conformations, $P(R,\Omega) = P(R|\Omega)P(\Omega)$.
This can be rewritten in energetic terms as 
\begin{equation}
E(R,\Omega) = -\ln P(R|\Omega) + E(\Omega)
\label{eq:12}
\end{equation}
where sequence-\emph{dependent} and sequence-\emph{independent} contributions
to the energy are separated. We assume that the sequence-independent term,
$E(\Omega)$, is defined by short-range interactions between the polypeptide
backbone, $\beta$-carbon and pseudo-atoms. At room temperature, van der Waals repulsions and
covalent bonding between atoms are extremely rigid interactions which contribute
to this energy. Another large contribution comes from hydrogen bonding, but the
magnitude of this interaction is vaguely understood. The sequence-dependent
part of the potential (the negative log-likelihood) can be approximated by the 
pair-wise interactions between side-chains, which make the largest contribution
to this term. In this work, these interactions are modelled by a G\={o}-type potential based 
on a `regularised' native contact map \cite{Podtelezhnikov:2009}, which contains lateral contacts
in parallel and anti-parallel $\beta$-sheets and contacts between residues $i$ and $i+3$ in $\alpha$-helices
\cite{leucine,leucine2}. Our model also includes a hydrophobic packing term; hydrophobic side chains coming into contact with hydrophobic or amphipathic side chains are rewarded with a decrease in energy \cite{LINUS}. The force constants for these side-chain interactions, as well as backbone hydrogen bonding, are optimised using a novel statistical machine learning technique \cite{Podtelezhnikov:2007il}.

Nested sampling is initialised with $K$ conformations, uniformly distributed over the space of dihedral angles (i.e.\ every $\phi_i,\psi_i \sim {U}[-180^\circ,180^\circ]$). To generate new sample points we use our implementation of an efficient Metropolis Monte Carlo (MMC) algorithm \cite{Podtelezhnikov:2008gm,Podtelezhnikov:2005so}, which relies on local Metropolis moves, as suggested in earlier studies \cite{Elofsson:1995sf}. In contrast to other programs that rely on local Metropolis moves in the space of dihedral angles, our sampler utilises local crankshaft rotations of rigid peptide bonds in Cartesian space. An important feature of our model is the elasticity of the $\alpha$-carbon valence geometry. With flexible $\alpha$-carbon valence angles, it becomes possible to use crankshaft moves inspired by earlier MMC studies of large-scale DNA properties. The amplitudes of proposed crankshaft rotations were chosen uniformly from $[-\alpha_0,\alpha_0]$ where, at every $2K$ nested sampling iterations, $\alpha_0$ (the maximum allowed proposed amplitude) was recalculated, attempting to keep the acceptance rate at 50\% (the trial MC moves used for this calculation were then ignored).

We ran simulations until $Z(\beta)$ converges for $\beta=1$ ($T = 25^{\circ}\mathrm{C}$), which implies that we have sampled from the thermodynamically accessible states for all temperatures smaller than $\beta$ (greater than $T$). The nested sampling algorithm marches left across the $x$-axis of Fig.~\ref{weight}. The step size is constant in $\log X$ and the larger $K$, the smaller the step size. For a given protein and $\beta$, we find that simulations terminate at approximately the same point on the $x$-axis (for protein G, with $\beta=1$, this is approximately $e^{-440}$). This implies that the total number of iterations is proportional to $K$, and the total number of MC moves is proportional to $mK$.  The results for protein G shown below are from a simulation with $K=20000$ and $m=15000$ which used 32 processors (Intel Xeon X5650), had 1.38x10$^{11}$ MC moves, and took about 22 hours.

\subsection{Energy Landscape Charts}

We use the algorithm recently introduced by P\'artay \emph{et al.} \cite{gab}, which uses the output of a nested sampling simulation to generate an \textit{energy landscape chart}, facilitating a qualitative understanding of potential energy surfaces. It has the advantage of showing the large scale features of the potential energy surface without requiring a large number of samples. 

The output of a nested sampling simulation is a sequence of sample points with decreasing energy. Each sample point (conformation), $\Omega_n$, represents $\omega_n = \alpha^{n-1}-\alpha^{n}$ of the phase space and has energy $E_n(\Omega_n)$.  A metric defining the distance between two conformations is required, and using this, a topological analysis of the sample points is performed. As the metric, we use the root mean square deviation (RMSD) of the backbone and side chain non-hydrogen atoms of a pair of conformations; that is, the sum of the Euclidean distances of corresponding atoms after the two conformations have been translated and rotated in space to minimise the overall distance.
 
A graph $\mathcal{G}$ is created with the sample points as nodes and arcs joining a sample to the $k$ nearest samples which have higher energy. In this work $k$ is chosen to be 15 throughout. We then start with an empty graph ($\mathcal{G}'$), adding nodes one at a time (starting with the lowest energy) to gradually rebuild $\mathcal{G}$.

Energy landscape charts are produced with energy on the vertical axis, and, at a given energy $E_n$, the width of the chart is proportional to the sum of the weights of the points below that energy (i.e.\ $\omega_n+\omega_{n+1}+\ldots$), that is, the available phase space volume in the prior space, contained below $E_n$.  On the horizontal scale, the chart is split into different basins corresponding to the disconnected subgraphs that exist when sample $n$ is added to $\mathcal{G}'$. The relative widths of the basins is given by the ratio of the sum of the weights of the sample points in the disconnected subgraphs.  The ordering of the basins horizontally is arbitrary.  Due to the rapid shrinking of the available phase space volume with decreasing energy, for better visualisation, a horizontal scaling is applied by an exponential function of the energy, similar to \cite{gab}. The energy landscape chart represents a potential energy landscape for the system. 

We also use a variant of the energy landscape charts where the width of the chart is proportional to the sum of the \textit{posterior} weights, $\chi_n = \omega_n \exp(-E_n\beta)/Z(\beta)$, i.e.\ ($\chi_n + \chi_{n+1} + ...$), at inverse temperature $\beta$.  Hence, the relative widths of the basins correspond to the probabilities of adopting a conformation from one basin or another.  These energy landscape charts, therefore, represent the energy landscape as it is `experienced' by the protein at inverse temperature $\beta$.  In the following, the two versions will be referred to as prior and posterior energy landscape charts, according to the weights used in the calculation of their basin widths.

\section{Results}
To validate the nested sampling procedure we simulated the folding of an isolated 16 residue polyalanine $\beta$-hairpin. We then conducted folding simulations on a number of small proteins which are commonly used for testing protein folding procedures: protein G (PDB code 1PGA),  the SH3 domain of Src tyrosine kinase  (PDB code 1SRL) and chymotrypsin inhibitor 2 (PDB code 2CI2).

\subsection{Isolated Polyalanine $\beta$-hairpin}
\begin{figure}[!ht]
\begin{center}
\includegraphics[width=3.25in]{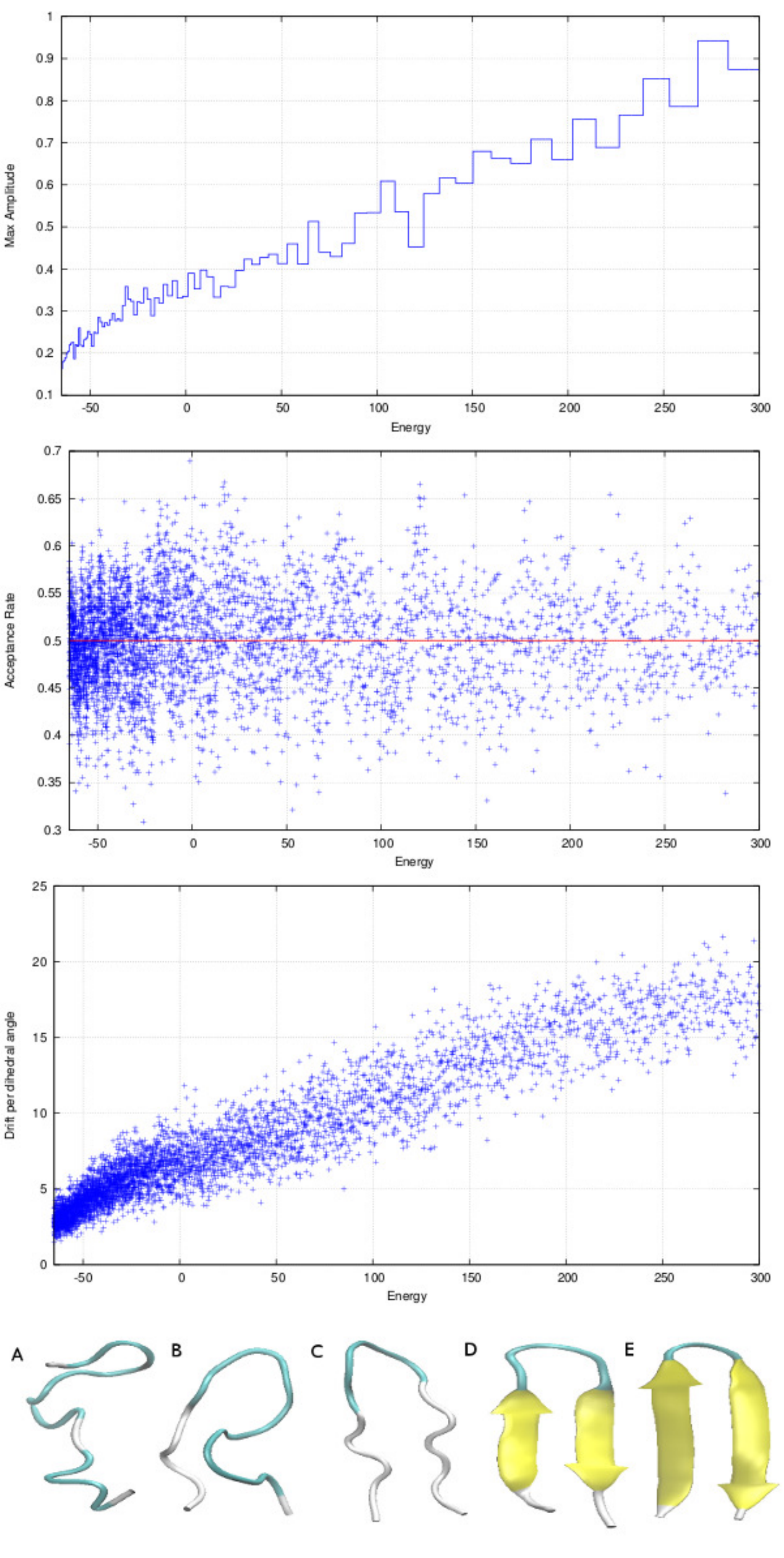}
\caption{Top graph: the maximum allowed amplitude of the crankshaft rotations $\alpha_0$ (in radians); Middle graph: the acceptance rate of the MC chains; Bottom graph: the drift per dihedral angle (the distance between the start and end conformations of a single MC chain). All with respect to the current energy threshold, for a 16 residue polyalanine $\beta$-hairpin; see text for more details.  Bottom: Five snapshots from a single nested sampling simulation of a $\beta$-hairpin with $K=1000$ and $m=2500$. The snapshots are equally spaced along the $\log(X)$ axis and have energies 3567, 190, 0, -46 and -66 (A-E). For comparison, the expectation of the internal energy at room temperature is -43. }
\label{amp}
\end{center}
\end{figure}

We used a G\={o}-like potential to simulate the folding of an isolated 16 residue polyalanine $\beta$-hairpin. Fig.~\ref{amp}~(bottom panel) shows a snapshot of five (equally spaced along the $\log(X)$ axis) conformations from a single simulation with $K=1000$, $m=2500$ (a total of 1.12x10$^8$ MC moves). At the beginning there is a rapid decrease in energy, moving from extended conformations (at first those with van der Waals collisions) to hairpin-like structures (A-C). The final part of the simulation moves through the exponentially small volume of the phase space containing hairpin-like structures, gradually decreasing in energy towards a fully formed hairpin (D-E).

We used the hairpin to check the behaviour of the nested sampling procedure: Fig.~\ref{amp}~(top panel) shows how $\alpha_0$ (the maximum proposed crankshaft rotation amplitude) varies with the energy threshold for a simulation with $K=1000$. As lower energy is reached, $\alpha_0$ is reduced to keep the acceptance rate near 0.5. Fig.~\ref{amp}~(second panel) shows the acceptance rate. Fig.~\ref{amp}~(third panel) shows the difference between the start and end points of a single MC chain, specifically the drift per dihedral angle, where the drift is the $L_2$-norm of the dihedral angles.

The protein model used stabilises room temperature secondary structure formation; it folds isolated helices and hairpins very effectively.  This is reflected in the energy landscape charts that consist of a single funnel (not shown).

\begin{figure}[!ht]
\begin{center}
\includegraphics[width=3.25in]{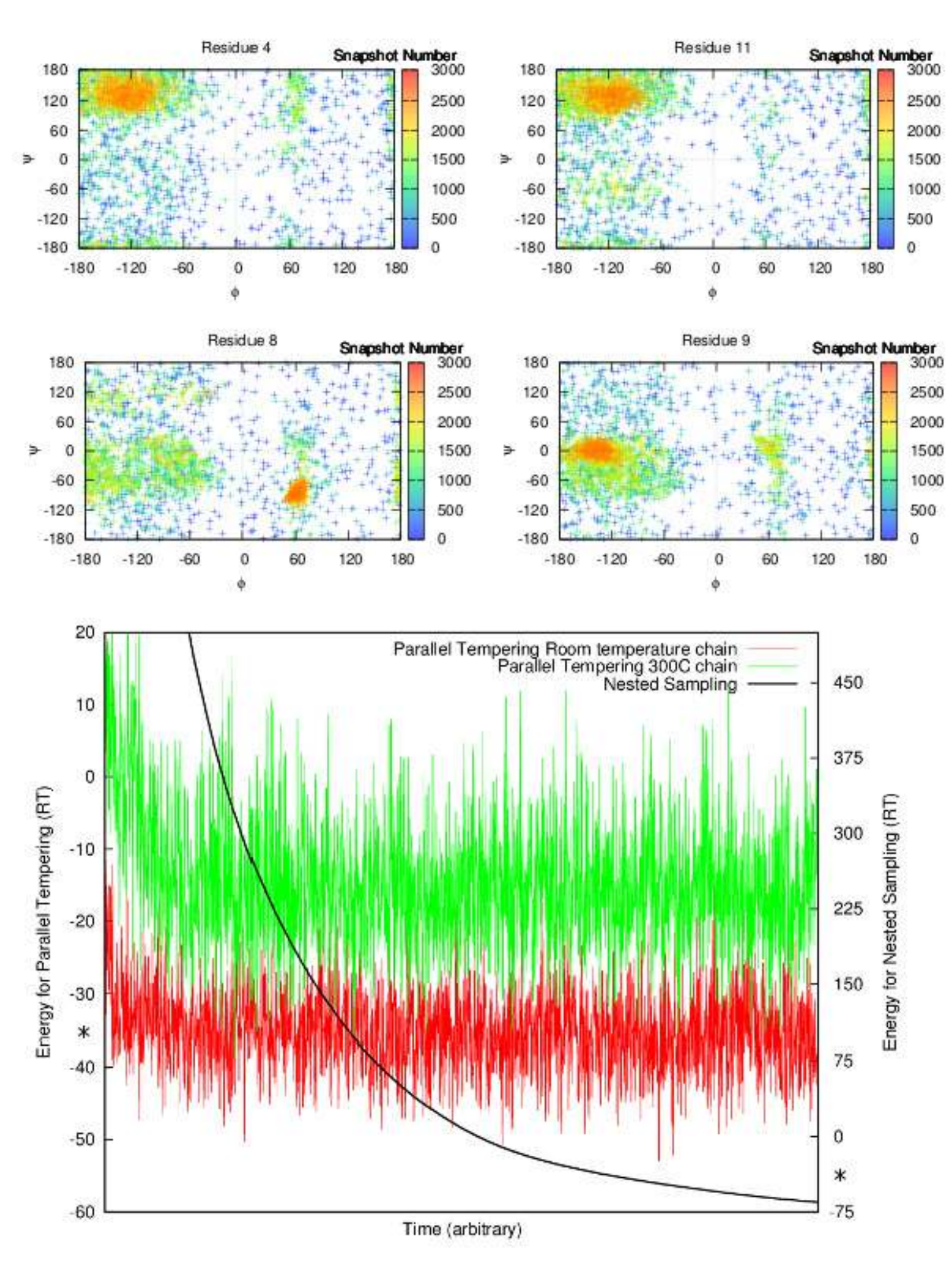}
\end{center}
\caption{Top: Dihedral angle evolution for residues 4, 8, 9 and 11 of the 16 residue polyalanine nested sampling simulation. In the later snapshots, residues 4 and 11 are distributed in the standard $\beta$-sheet region of the Ramachandran plot. Residues 8 and 9 contain the turn of the polypeptide.  The dihedral angles of the turn residues, 8 $(60 \pm 15,-90 \pm 30)$ and 9 $(-150 \pm 30,0 \pm 30)$, are closest to the values of type II' turn ($(60,-120)$ and $(-80,0)$) \cite{hairpintypes}. Bottom: Energy v Time graph for nested sampling (black, right hand axis) and two of the chains from a parallel tempering simulation (red, room temperature; green, 300$^{\circ}$C; both left hand axis). On both vertical axes a star marks the expected thermodynamic energy at room temperature. }
\label{hairpinram}
\end{figure}

\Red{Fig.~\ref{hairpinram} (top) shows the time evolution of the dihedral angles of 4 residues of the 16 residue polyalanine. The formation of the hairpin can be clearly seen. For example, the dihedral angles of the residues in the strands 4 and 11 converge to the standard $\beta$-sheet area of the Ramachandran plot. The G\={o}-like potential used was designed for a hairpin with a two residue turn, and this is found to be the case. The dihedral angles of the turn residues 8 $(60 \pm 15,-90 \pm 30)$ and 9 $(-150 \pm 30,0 \pm 30)$, are closest to the values of a type II' turn ($(60,-120)$ and $(-80,0)$) \cite{hairpintypes}. Fig.~\ref{hairpinram} (bottom) shows the energy of the snapshots (right hand axis) for nested sampling plotted against time. The montonic decrease of the energy over a very large energy range allows us to view the formation of the hairpin.} 

\Red{Due to the nature of the model used, the folding pathway of the hairpin is relatively simple to sample, and parallel tempering can also successfully fold the hairpin. However, in this case, we need a very large temperature range to explore the whole parameter space and view the folding pathway in its entirety. For example, Fig.~\ref{hairpinram} (bottom) shows the energy of 2 of the parallel tempering chains; room temperature (red) and 300${^\circ}$C (green). For real proteins, which have more complicated energy landscapes and possibly high energy barriers, it is difficult to know the temperature range required for parallel tempering to explore the entire parameter space and not be `trapped' in a particular basin. Nested sampling, with its top down, temperature independent approach, does not suffer from this problem.}

\begin{figure}[!ht]
\begin{center}
\includegraphics[width=3.25in]{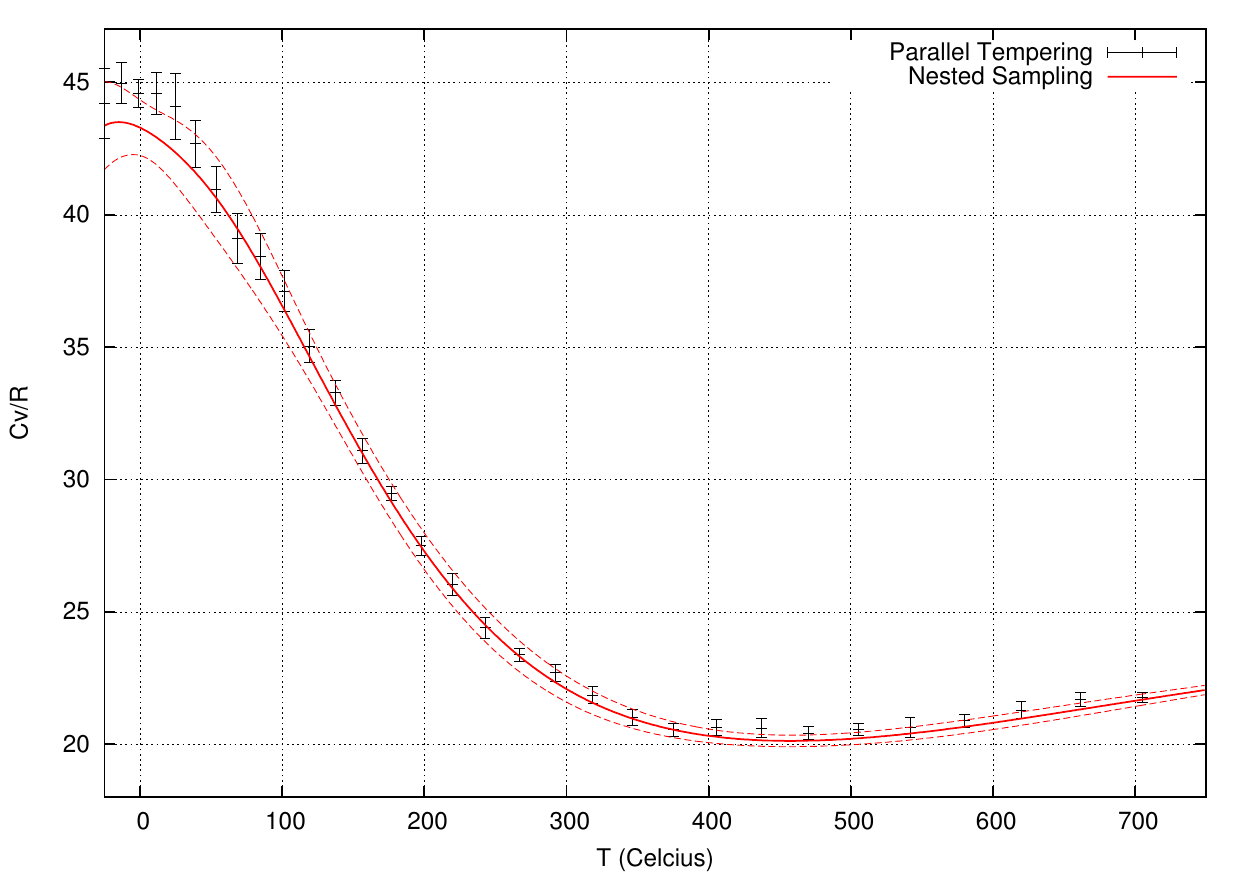}
\caption{The heat capacity curve for the 16 residue polyalanine. The nested sampling simulations (solid line) use 1.3x10$^9$ MC moves, with error lines denoting 2 standard errors from the mean. The parallel tempering uses the same number of MC moves again with error bars showing 2 standard errors from the mean. }
\label{HC}
\end{center}
\end{figure}

\Red{Another of} the advantages of nested sampling is that simulations are temperature independent, and hence can provide estimates of thermodynamic variables at any temperature. Fig.~\ref{HC} shows the heat capacity ($C_v$) curve for the 16 residue polyalanine. The curves were calculated using nested sampling (converged down to $-25^{\circ}\mathrm{C}$, so that the $C_v$ curve does not stop abruptly at room temperature), and parallel tempering. The solid line is calculated using 10 nested sampling simulations each with 1.3x10$^9$ MC moves. The dashed lines show twice the standard error. The parallel tempering curve shows the heat capacity using 10 parallel tempering simulations (again each with 1.3x10$^9$ MC moves) with error bars showing twice the standard error. For parallel tempering, the heat capacity is only calculated for discrete temperatures and a procedure such as Boltzmann reweighting \cite{reweight} is needed to calculate the continuous curve.
 
There appears to be good agreement between the methods. Previous results have found nested sampling to be more efficient at calculating the heat capacity curves \cite{gab}. In this example, we found nested sampling to be of similar efficiency to parallel tempering. We believe this to be because, unlike the system presented in \cite{gab}, our phase transition (from coil to hairpin) occurs over a very large energy (and hence temperature) range from which parallel tempering can successfully sample. 

\subsection{Protein G}
\begin{figure}[!ht]
\begin{center}
\includegraphics[width=3.25in]{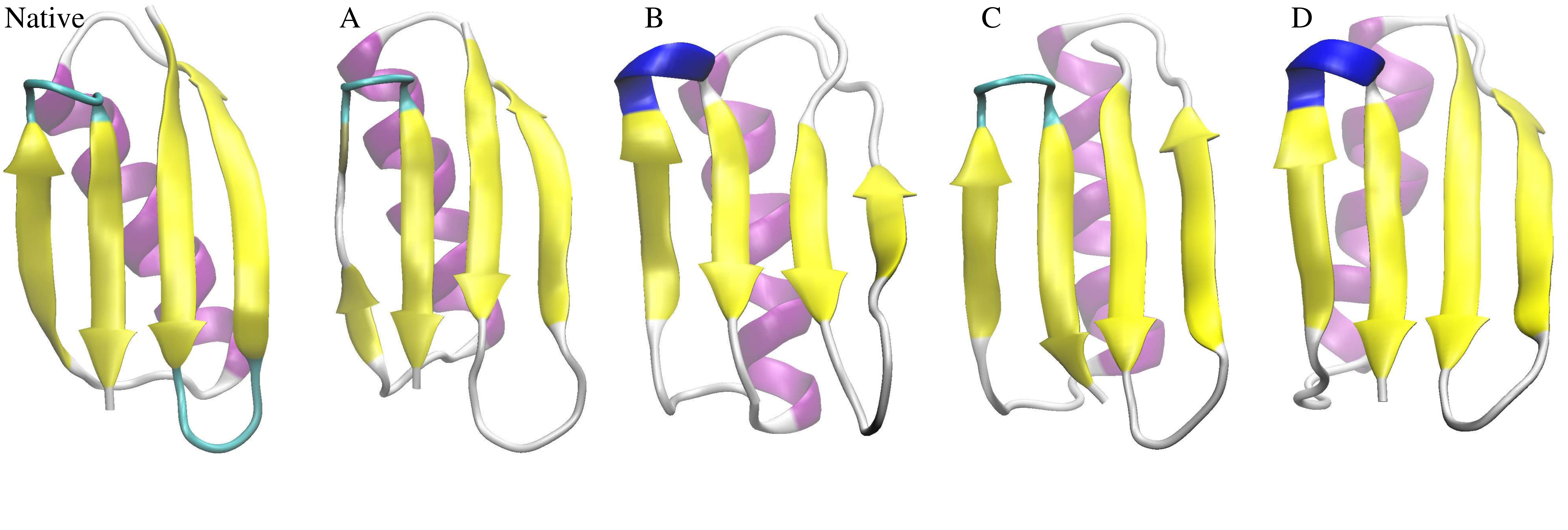}
\caption{The native (crystal) structure of protein G (left) with a sample of conformations accessible at room temperature from a simulation with $K=20000$ and $m=15000$. All figures of protein G in this paper have been oriented so that the first $\beta$-strand is the second strand from the right and the N-terminal residue is at the top. The thermodynamic energy at room temperature, estimated from the simulation, is -190 and conformers A,B,C and D have have energies -189, -190, -191 and -190, respectively. The backbone RMSDs from the crystal structure are $1.93\,\text{\AA}$, $2.96\,\text{\AA}$, $3.97\,\text{\AA}$ and $5.22\,\text{\AA}$, respectively.  The angle between the helix projected onto the sheet and the first $\beta$-strand is $17.9^\circ$, $8.6^\circ$, $-4.7^\circ$ and $-15.1^\circ$, respectively, compared to $21.9^\circ$ of the crystal structure.
 }
\label{fourg}
\end{center}
\end{figure}

Protein G  is a 56-residue protein consisting of an anti-parallel four-stranded $\beta$-sheet and an $\alpha$-helix, with a $\beta$-Grasp (ubiquitin-like) fold, which has been extensively studied  by a variety of folding simulation techniques \cite{karanicolas2002origins, Kolinski:2004az, Sheinerman:1998zg, Shimada:2002hw}. Its native structure is shown on the left of Fig.~\ref{fourg}.  All figures of protein G in this paper have been oriented so that the first $\beta$-strand is the second strand from the right and the N-terminal residue is at the top. 

As described above, the nested sampling procedure can be used to estimate the thermodynamic energy of the system at any temperature, using equation \ref{eq:thd}. For protein G, at room temperature ($\beta = 1.0$), the thermodynamic energy is -190 units. Fig.~\ref{fourg} shows a sample of four room temperature, thermodynamically accessible conformations found by a single nested sampling simulation with $K=20000$ and $m=15000$. The conformers have energies -189, -190, -191 and -190, respectively, with backbone RMSDs (from the crystal structure) of $1.93\,\text{\AA}$, $2.96\,\text{\AA}$, $3.97\,\text{\AA}$ and $5.22\,\text{\AA}$ respectively.  The estimated value of the backbone RMSD at $\beta=1$, calculated using equation~\ref{eq:thd}, is $\mathbb{E} \left(\mathrm{RMSD} | \beta = 1 \right) = 3.21\,\text{\AA}$.

Conformers A--D have the correct backbone topology, close to the native structure, but there is a reasonable amount of variation in the orientation of the helix with respect to the $\beta$-sheet at this temperature.  It is important to remember that protein structures are intrinsically flexible \cite{HenK07,ThoLRJK01,WelMHT05}, and the crystal structure (\mbox{1PGA.pdb}) is only one member of an ensemble of conformations that the protein may explore. In Appendix C we demonstrate that flexible motion of protein G allows a substantial reorientiation of the axis of the helix with respect to the sheet.  Conformers A--D in Fig.~\ref{fourg}, which differ from the native state principally in the orientation of the helix relative to the sheet,  may therefore be more representative of the native state than the RMSD alone suggests.

\begin{figure}[!ht]
\begin{center}
\includegraphics[width=3.25in]{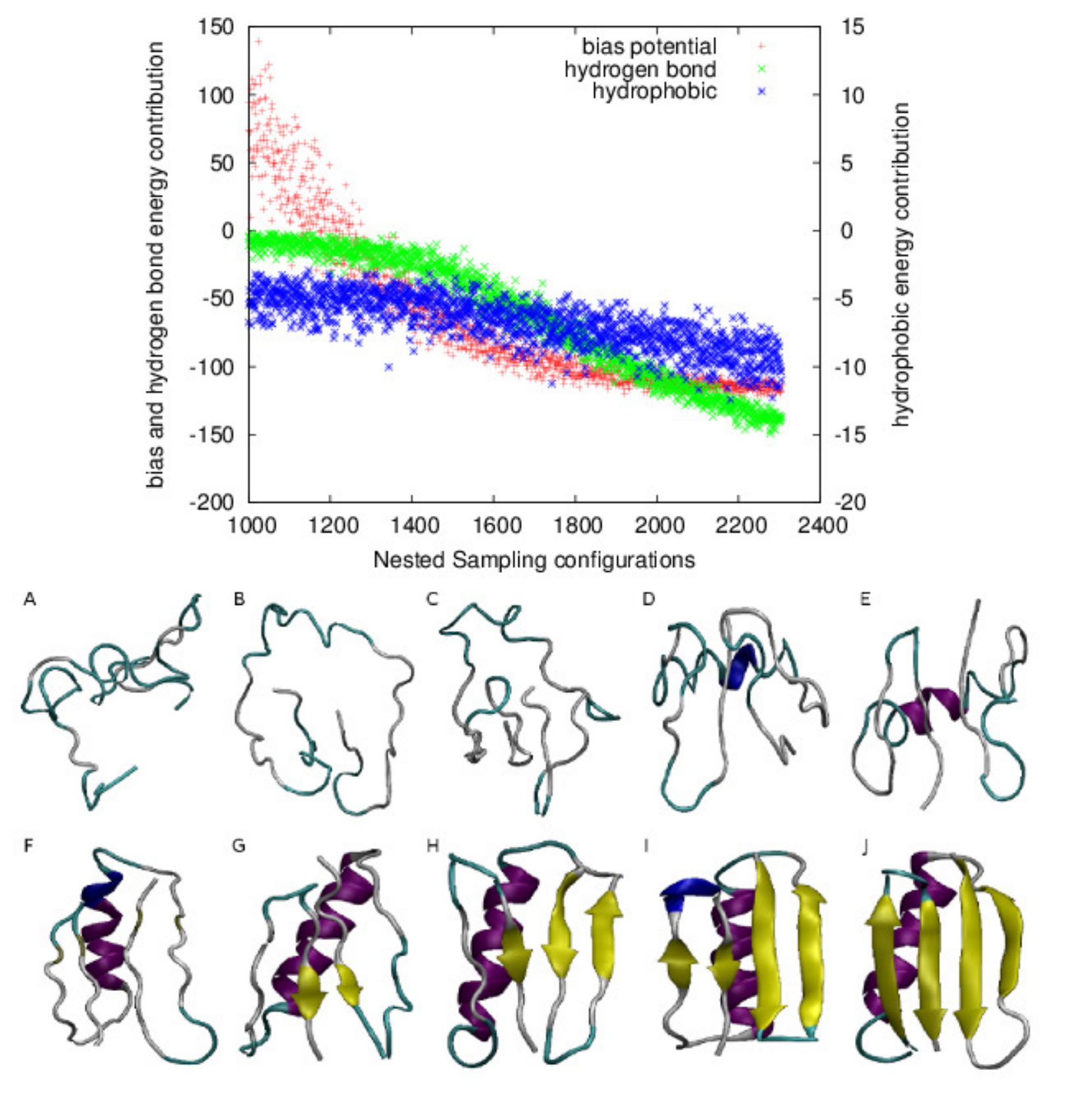}
\caption{Top: Energy contributions of the G\={o}-type bias potential (red), hydrogen bonds (green) and hydrophobic interactions (blue) in the second half of a nested sampling simulation of protein G with $K=20000$ and $m=15000$.  Units of energy are in $RT$ corresponding to temperature.  Note the different scale on the vertical axes.  Bottom:  Ten conformations of protein G from the same simulation, in order of decreasing energy. The colours are: purple; $\alpha$-helix, dark blue; $3_{10}$-helix, yellow; $\beta$-strand, cyan; turn, white; coil. }
\label{sequence}
\end{center}
\end{figure}

\begin{figure}[!ht]
\begin{center}
\includegraphics[width=3.25in]{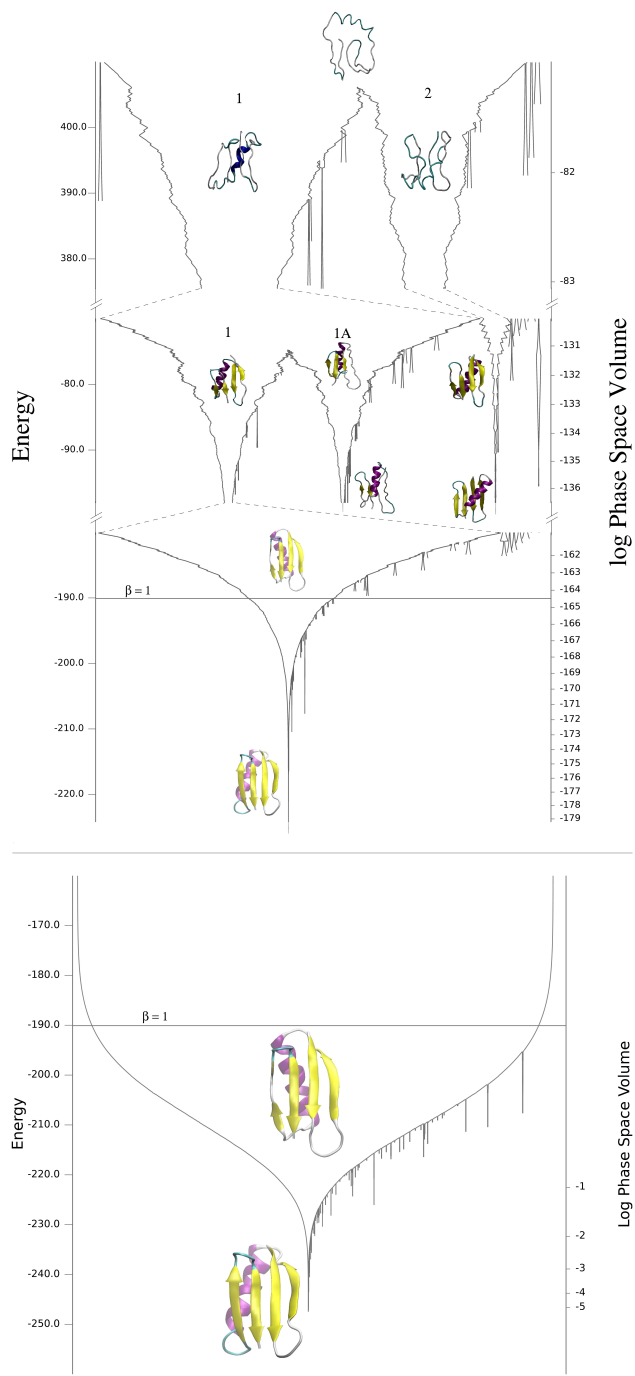}
\caption{Top: prior (potential) energy landscape chart, Bottom: posterior energy landscape chart at $\beta=1$, for a nested sampling simulation of protein G using $K=20000$ and $m=15000$.  On the left axis the energy is shown in units of $RT$, and the width of the chart is proportional to the sum of the prior (Top) and posterior (Bottom) weights of the nested sampling points below the given energy level (shown on the right axis). The prior energy landscape chart shows the potential energy surface and the posterior energy landscape chart shows, for a given temperature, the probabilities of finding conformations from the different basins. At $\beta=1$ (room temperature) only funnel 1 is accessible. The scaling function used for the prior energy landscape chart is $\exp(-f E)$ with $f$ being 0.1, 0.4 and 0.7 on the top, middle and bottom panels of the prior energy landscape chart, respectively.  Example conformers from the main basins, at various energy levels, are shown on the charts.
}
\label{elc1}
\end{center}
\end{figure}

The first half of the nested sampling simulation is spent exploring high energy conformations with no noticeable secondary structure and often steric hindrances.  In the second half of the simulation, once the long range quadratic bias potential has pulled the secondary structure elements close together, the short range hydrogen bond interaction contributions increase to dominate the bias potential contributions, having a steeper gradient in the last third of the simulation (Fig.~\ref{sequence}~(top)).  The short range hydrophobic interaction contributions are the smallest, but nevertheless not negligible; they ensure the correct packing of the hydrophobic and amphipathic side chains at confomations available at room temperature (see below and Fig.~\ref{elc1}).
Fig.~\ref{sequence}~(bottom) shows a sequence of 10 conformers in order of decreasing energy.  These conformers come from the deepest basin of the energy landscape chart of a simulation (higher energy conformers come from the part of the energy landscape chart which contains the deepest basin). The sequence illustrates how the secondary and tertiary structure \Red{accrete} in the course of a simulation, \Red{capturing the essence of the hierarchical folding model}. The sequence is not, however, a single folding pathway, in the sense of a molecular dynamics trajectory; there are many conformations in the active set in the same basin. It is possible, though, that Fig.~\ref{sequence}~(bottom) represents a plausible sequence of events leading to the native structure.

Energy landscape charts using the prior and posterior weights for a nested sampling simulation of protein G using $K=20000$ and $m=15000$, calculated using a connectivity number $k=15$, are shown in Fig.~\ref{elc1}. The volume scale on the right hand axis shows the proportion of the prior and posterior phase space volume available below the given energy level.  The width of the chart uses this scale.  Basins which contain less than 1/1000th of the probability mass at the point of splitting are not shown on the diagram. Conformers have been placed on the chart to provide examples of the samples found in different places of the chart. 

Topologically, for energy above 405 units, there is one main basin containing virtually all of the samples. There is little structure in the samples, as shown by the conformers on the top panel of the chart. However, at energy 405 units, the phase space splits into two main funnels: one with the helix forming on the correct side of the sheet (funnel 1) and one with it forming on the incorrect side (funnel 2).  Funnel 1 further splits at energy -75 units, corresponding to conformations where the hydrophobic residues are in the interior of the protein (funnel 1) or on the surface of the protein (funnel 1A).  At room temperature (the expected energy corresponding to $\beta=1$ is marked by a horizontal line on the bottom panel of the chart), the phase space volume of both funnels 1A and 2 are less than 1/1000th of the main funnel and hence the posteror energy landscape chart consists of a single funnel. 
The inaccessibility of funnel 1A at room temperature indicates the importance of hydrophobic interactions.
\begin{figure}[!ht]
\begin{center}
\includegraphics[width=3.25in]{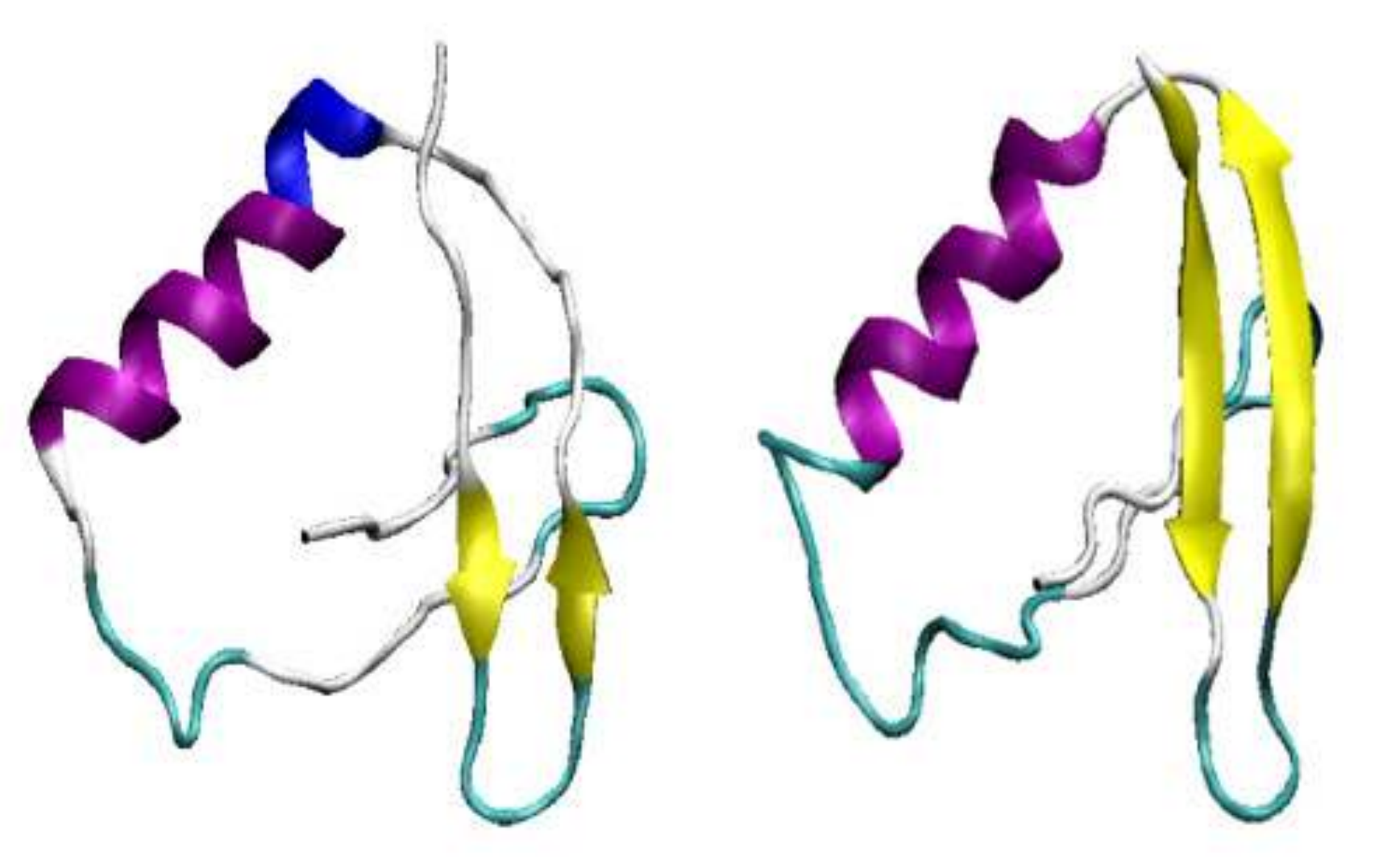}
\end{center}
\caption{The topological analysis places similar conformations in the same basin. For example, these two conformations (which both have the 3rd and 4th $\beta$-strands aligned incorrectly) are placed in the same basin.
} 
\label{elc2}
\end{figure}

Fig.~\ref{elc2} shows two conformers which are placed in the same small basin, branching off the right hand funnel. The conformer on the left has higher energy than the one on the right. These conformers are very similar, and demonstrate that the topological analysis shows how meta-stable conformations are formed. The pathway to these states would be obtained by considering conformers found in the same basin. 
\clearpage

\section{Discussion}
It is interesting to consider how the energy landscape charts vary from simulation to simulation. Topologically, we always find two main funnels in the protein G simulations (funnels 1 and 2 on Fig.~\ref{elc1}), corresponding to the packing of the helix on the two sides of the helix, and the dominant mode with the native-like backbone topology (funnel 1) splits again at a lower energy level to two funnels, corresponding to the hydrophobic residues being in the interior (funnel 1) or on the surface (funnel 1A) of the protein.  The energy at which funnels 1 and 2 split varies significantly between simulations, from 220 to 580 energy units.  
This is probably due to the fact that the RMSD metric is an imprecise way of comparing wildly different conformations.
The energy where funnels 1 and 1A split has a much smaller variation, $-75$ to $-55$ energy units.  This trend in the variation of splitting energies was also observed in the nested sampling simulations of the other modeled proteins. Metrics other than the RMSD might improve the reproducibility of energy landscape charts and would be worthy of investigation.

The relative basin widths of energy landscape charts depend on the size of the nested sampling active set, $K$.
In general, $K$ determines the resolution of exploration.  When converging the evidence at lower temperatures, a larger value of $K$ is required.  This is because at every splitting of the likelihood function, the probability of exploring the dominant mode decreases according to~equation~\ref{dominant-mode-prob}.
At high energies, the accessible conformational space is connected, and the MMC procedure explores the space effectively.  As the energy lowers, the accessible conformational space becomes increasingly disconnected.  Since the MMC procedure cannot jump between disconnected components of the conformational space, an increasingly large set of active points is required to sample effectively.  As the posterior mass is concentrated at lower energies for lower temperatures, $K$ behaves as an effective minimum temperature.  Using too small an active set for a given temperature causes large variation between different nested sampling simulations; for example, the estimates for the evidence and the relative widths of the funnels of energy landscape charts.

In the protein G simulations, we find that $K=20000$ is large enough to produce simulation independent charts for temperatures near $\beta=1$.
When using, for example, $K=2500$, which is too small for sampling the posterior distribution at $\beta=1$, we find that the active set becomes extremely homogenous and the simulation is, in effect, exploring just one tiny basin in one of the main funnels, by making smaller and smaller crankshaft rotations.  Hence, we find a single room temperature accessible conformation, as opposed to the wide selection that is found when $K=20000$.

\Red{The magnitude of $m$ relative to $K$ is problem specific. It has been suggested that for probability distributions which lack a large number of modes, it is optimal to set $K$ small and use a large $m$ (the cost is proportional to $mK$) \cite{potts}. For Protein G, we find the energy landscape is so complex that we need a large $K$ in order to explore all the funnels simultaneously, and a large $m$ to ensure the active set remains heterogeneous, and we therefore choose $m$ and $K$ to have the same order of magnitude. Incorporating non-local flexibile motions \cite{JimFRW11} into our MMC procedure may allow a decrease in $m$ without losing heterogenity and this is a focus of future work. If this proves to be the case we would choose to increase $K$ relative to $m$.}

 In our previous work \cite{Podtelezhnikov:2009}, \Red{using MMC} with parallel tempering to simulate the folding of protein G with a simpler model (no $\gamma$-atoms and hydrophobic interactions were included in this model), the lowest energy structures obtained were similar to those shown at the bottom of funnel 2 of Fig.~\ref{elc1}, with the helix packed on the incorrect side of the sheet and a backbone RMSD of $8.6\,\text{\AA}$ from the crystal structure. This demonstrates the difficulty of using parallel tempering or simulated annealing to reconstruct the native structure, when the energy landscape exhibits two main funnels separated by a large energy barrier.  If the annealing proceeds down the `incorrect' funnel it will be nearly impossible for it to climb back out and down into the `correct' funnel. The reason for the double funnel is the symmetry of the protein G topology with respect to the G\={o}-type bias potential,
  which is the predominant factor at the beginning of the simulation.  The further splitting of the main funnel into funnels 1 and 1A (Fig.~\ref{elc1}) is also due to the nature of the G\={o}-type bias potential.  This applies a quadratic potential on the C$_\beta$ atom contacts, which does not restrict the hydrogen bond pattern between the individual strands; at high energies, both conformations (with the hydrophobic residues of the $\beta$-sheet being in the interior or on the surface of the protein) are similarly likely to be adopted.  However, other energy and entropy contributions due to the presence of side chains (e.g. hydrophobic interactions and steric clashes) ensure that only conformations with the native-like topology are accessible at room temperature.
This way, energy landscape charts also reflect the nature of the protein model and force field used.
For example, the energy landscape charts for chymotrypsin inhibitor 2, which differs in topology from protein G, but also possesses  a similar symmetry with regard to the packing of the $\alpha$-helix against the $\beta$-sheet, also exhibit this double funnel (see Appendix B).
It would be interesting to compare energy landscape charts of nested sampling simulations using other protein models and force fields, for example, all-atom representations, \Red{and this will be a focus of future work}.

\section{Conclusion}
This paper has described the parallelisation of the nested sampling algorithm, and its application to the problem of protein folding in a force field of empirical potentials that were designed to stabilise 
secondary structure elements in room-temperature simulations. The output of the nested sampling algorithm can be used to produce energy landscape charts, which give a high level description of the potential energy surface for the protein folding simulations. These charts provide qualitative insights into both the folding process and the nature of the model and force field used. The topology of the protein molecule emerges as a major determinant of the shape of the energy landscape, as has been noted by other authors \cite{karanicolas2002origins}. The energy landscape chart for protein G exhibits a double funnel with a large energy barrier, a potential energy surface which parallel tempering struggles to explore fully. The nested sampling algorithm also provides an efficient way to calculate free energies and the expectation value of thermodynamic observables at any temperature, through a simple post-processing of the output.

\section{Acknowledgements}
We thank  G\'{a}bor Cs\'{a}nyi, Konrad Dabrowski, L\'{i}via P\'{a}rtay, Alexei Podtelezhnikov and John Skilling for helpful discussions. We acknowledge support from the Leverhulme Trust  (grant F/00 215/BL) (NB, CV and DLW) and the EPSRC (grant EP/GO21163/1) (DLW). SAW acknowledges support from the Leverhulme Trust (Early Career Fellowship).

\bibliography{nested}

\appendix

\section{The Protein Model}
We modelled the polypeptide as a chain of peptide groups elastically connected at the 
C$_\alpha$ atoms, with the valence angles constrained to $111.5^{\circ} \pm
2.8^{\circ}$. The positions of all backbone and C$_\beta$ atoms, including hydrogen, were
specified by the orientations of the peptide bonds. We fixed the peptide bond
lengths and angles at standard values \cite{Engh:1991if, Engh:2001xu,
Brunger:1992jx}. The distance between C$_\alpha$ atoms separated by \textit{trans}
peptide bonds was fixed at 3.8~\AA. The C$_\beta$ positions were stipulated by
the tetrahedral geometry of the C$_\alpha$ atoms and corresponded to
\textsc{l}-amino acids. Most of the conformational variability of polypeptides
comes from relatively free rotation around N--C$_\alpha$ and C$_\alpha$--C
bonds characterised, respectively, by dihedral angles $\phi$ and $\psi$ (Fig.~1 in \cite{Podtelezhnikov:2005so}). These
rotations are least restricted in glycine that lacks C$_\beta$. The dihedral
angles $\phi$ in proline were elastically constrained to $-60^{\circ} \pm
7^{\circ}$ by covalent bonding \cite{Ho:2005qd}. We introduced a harmonic
potential $E_i^B$ to impose these and other elastic constraints.  
A more detailed description of the model is given in our previous work \cite{Podtelezhnikov:2005so}.

In this work, we represented other side chain atoms by one, or in the case of branched side chains, two pseudo-atoms, following \cite{LINUS}. The side chain dihedral angles $\chi$ were permitted to vary, and take the values $\{\pm 60^{\circ}, 180^{\circ}\}$, or in the case of proline $\{\pm 30^{\circ}\}$, with probabilities dependent on residue type, with values corresponding to the distribution of the $\chi$ angles in the same ASTRAL PDB database~\cite{astral} that was used in \cite{Podtelezhnikov:2005so}, and here, to learn the potential parameters by \Red{a statistical machine learning procedure,} contrastive divergence \cite{Carreira-Perpinan:2005bq}.


We modelled van der Waals repulsions so that there is a
prohibitively large energetic cost of overlaps between atoms. We used values of
 atomic radii close to a lower limit of the range found in the
literature \cite{Ramachandran:1963yu, Hopfinger:1973jt,
Word:1999rm,Pappu:2000ty}: $r(\mbox{C}_\alpha) = r(\mbox{C}_\beta) =
1.57~\mbox{\AA}$, $r(\mbox{C}) = 1.42~\mbox{\AA}$, $r(\mbox{O}) =
1.29~\mbox{\AA}$, $r(\mbox{N}) = 1.29~\mbox{\AA}$. We adopted values of the contact radii for the pseudo-atoms from \cite{LINUS}.

Hydrogen bonding is a major polar interaction between NH and CO groups of
polypeptide backbone. Based on surveys of the Protein Data Bank (PDB)
\cite{Berman:2000qr}, important reviews of hydrogen bonding in globular
proteins have formulated the basics of the current understanding of hydrogen
bond geometry and networking \cite{Baker:1984qr, Savage:1993fb, Stickle:1992it,
McDonald:1994ca}. We considered the hydrogen bond formed when three distance
and angular conditions were satisfied: $r(\mbox{O, H}) < \delta$, $\angle
\mbox{OHN} > \Theta$, and $\angle \mbox{COH} > \Psi$, where $r(\mbox{O, H})$ is
the distance between oxygen and hydrogen, and symbol $\angle$ denotes the angle
between the three atoms (see Fig.~1A in \cite{Podtelezhnikov:2007il} ). The lower bound on the
separation between the atoms ($r(\mbox{O, H}) > 1.8~\mbox{\AA}$) was implicitly
set by the hard-sphere collision between oxygen and nitrogen. We used the same
hydrogen bond potential regardless of the secondary structure adopted by the
peptide backbone. The energy of the hydrogen bond (Fig.~1B in \cite{Podtelezhnikov:2007il} ) was
described in \cite{Podtelezhnikov:2007il} by a square-well potential,
\begin{equation}
E_{ij}^{HB} = -n_hH
\label{eq:13}
\end{equation}
where $H$ is the strength of each hydrogen bond, and $n_h$ is the number of
hydrogen bonds between the amino acids $i$ and $j$. The strength of
the hydrogen bonds, $H$, as well as the three cutoff parameters, $\delta$,
$\Theta$, $\Psi$ was determined by a machine learning procedure, contrastive divergence \cite{Carreira-Perpinan:2005bq}. We found that softening the hard cutoffs $\{\delta, \Psi , \Theta \}$ improved the results, and hence we used a steep continuous approximation to the square well.

We modelled hydrophobic interactions in a manner consistent with \cite{LINUS}.  The hydrophobic interaction contribution between hydrophobic atoms $A$ and $B$ of amino acids $i$ and $j$ ($|i-j| \ge 2$) is 
\begin{equation}
E_{AB}^{hyd} = \left\{
\begin{array}{ll}
f k_h & r_{AB} < r_{cut,AB} \\\\
f k_h \left( \frac{r_{AB}-r_{cut,AB}}{\Delta} \right) & r_{cut,AB} \le r_{AB} < r_{cut,AB} + \Delta
\end{array}
\right.
\end{equation}
where $k_h$ is a constant parameter proportional to the Kauzmann parameter~\cite{kauzmann}, the cutoff distance $r_{cut,AB}$ is the sum of the vdW radii of atoms $A$ and $B$ listed in~\cite{LINUS}, $\Delta=2.8\ \text{\AA}$ is a smoothing range beyond the cutoff distance, and the multiplicative factor $f$ takes the value 2, if both amino acids are hydrophobic, 1, if one is hydrophobic and the other one is amphipathic, and 0, if neither are hydrophobic.  Hydrophobic amino acids are cysteine, isoleucine, leucine, methionine, phenylalaninne, tryptophan and valine; amphipathic residues are alanine, histidine, threonine and tyrosine.

The sequence-\emph{dependent} part of the potential (the negative
log-likelihood) was approximated in our model by pair-wise interactions between
side-chains, as described in~\cite{Podtelezhnikov:2009}. Our main focus was on the resulting effect of these interactions
and how they stabilise secondary structural elements. We did not consider the
detailed physical nature of these forces, or how they depend on the amino acid
types. We introduced these interactions between the polypeptide side chains as
an effective G\={o}-type potential \cite{Go:1983bk} dependent on the distance
between C$_\beta$ atoms,
\begin{equation}
E_{ij}^{SC} = \kappa C_{ij} (r_{ij} - r ) ^ 2
\label{eq:14}
\end{equation}
where $r_{ij}$ is a distance between non-adjacent C$_\beta$ atoms, $|i-j| > 1$; $r$ a constant
and $\kappa$ is a force constant. In \cite{Podtelezhnikov:2009} we introduced a ``regularised
contact map", $C_{ij}$. In this binary matrix, two types of contacts were
defined in the context of protein secondary structure. First, only lateral
contacts in the parallel and anti-parallel $\beta$-sheets were indicated by
1's. Second, the contacts between amino acids $i$ and $i + 3$ in
$\alpha$-helices were also represented by 1's. These contacts typically have the closest 
C$_\beta$--C$_\beta$ distance among
non-adjacent contacts in native proteins. The force constants and $r$ depend on the
secondary structure type, introducing positive $\kappa_\alpha$ 
$\kappa_\beta$, $r_\alpha$ and $r_\beta$. Non-adjacent contacts in secondary structural elements were,
therefore, stabilised by attracting potentials.

We also modelled interactions between sequential residues. This interaction was
defined by the mutual orientation of adjacent residues that are involved in
secondary structural elements,
\begin{equation}
E_{i,i+1}^{SC} = \eta \cos \gamma_{i,i+1}
\label{eq:15}
\end{equation}
where $\gamma_{i,i+1}$ is the dihedral angle
N$_i$--C$_{\alpha,i}$--C$_{\alpha,i+1}$--C$_{i+1}$ between the adjacent residues. The
purpose of this interaction is to bias the conformation towards the naturally
occurring orientations of residues in secondary structural elements. In
$\alpha$-helices, adjacent residues adopt a conformation with positive
$\cos \gamma$. In $\beta$-sheets, $\cos \gamma$ is negative. We, therefore, used
two values of the force constant: negative $\eta_\alpha$ and positive
$\eta_\beta$.

As in~\cite{Podtelezhnikov:2009}, all parameters were determined by a \Red{statistical} machine learning procedure, contrastive divergence \cite{Carreira-Perpinan:2005bq} and in this work $\delta=2.06,\: -\cos \Theta = 0.89,\: -\cos \Psi = 0.766, \: H = 4.35,\: \eta_\beta = 3.5,\: \eta_\alpha = -4.9,\: \kappa_\alpha = 3.3,\: \kappa_\beta = 1.2,\: r_\alpha = 5.66,\: r_\beta = 5.35$ and $k_h=0.08$, where the unit of energy is $RT$ at room temperature. \Red{With the improved model and force field described in this paper, contrastive divergence provided good parameters without the need of further adjustments, as had been the case in~\cite{Podtelezhnikov:2009}. }

To summarise, the total energy of a polypeptide chain with conformation
$\Omega$ was calculated as follows
\begin{equation}
E(R,\Omega) = \sum_{i=1}^N E_i^B +
\sum_{i=1}^N \sum_{j=1}^i (E_{ij}^{vdW} + E_{ij}^{HB} + E_{ij}^{SC} + E_{ij}^{hyd})
\label{eq:16}
\end{equation}
where we consider harmonic valence elasticity, $E_i^B$, van der
Waals repulsions, $E_{ij}^{vdW}$, hydrogen bonding,
$E_{ij}^{HB}$ and hydrophobic packing, $E_{ij}^{hyd}$. The valence elasticity, van der Waals repulsions, and hydrogen bonding that contribute to this potential have a clear physical meaning and are
analogous to traditional \emph{ab initio} approaches. The side-chain
interactions, $E_{ij}^{SC}$ in this model were introduced as a long-range
quadratic G\={o}-type potential based on the contact map and secondary
structure assignment. This pseudo-potential had two purposes: it was needed to
stabilise the secondary structural elements, and to provide a biasing force that
allows reconstruction of the backbone conformation in the course of Metropolis
Monte Carlo simulations \cite{Podtelezhnikov:2005so,Podtelezhnikov:2009}.

\section{Additional Results}
\subsection*{Src Tyrosine Kinase SH3 Domain.}
\label{1srl}
\begin{figure}[!tpb]
\begin{center}
\centerline{\includegraphics[width=.7\textwidth]{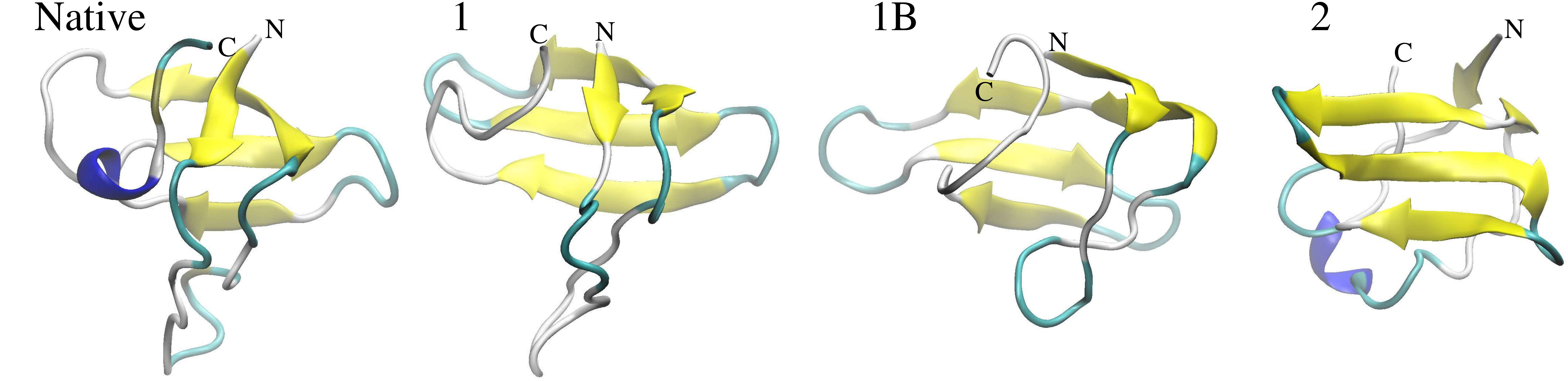}}
\caption{The native (crystal) structure of the SH3 Domain, with conformers from the two main funnels accessible at room temperature from a simulation with $K=15000$ and $m=15000$. Conformers 1 and 1B have the correct backbone topology, unlike conformer 2.  The backbone RMSDs are 4.89\,\AA, 5.28\,\AA\ and 11.51\,\AA, respectively.  The N and C termini are marked; see the text for more details.}
\label{fourone}
\end{center}
\end{figure}

Src Tyrosine Kinase SH3 Domain is a 56-residue protein, comprising a 5-stranded $\beta$-barrel. 
The last strand is interrupted by a single turn of a $3_{10}$-helix, which was not included in the `regularised' contact map used to define the native state in these simulations.
The native (crystal) structure is shown in Fig.~\ref{fourone}.

\begin{figure}[!tpb]
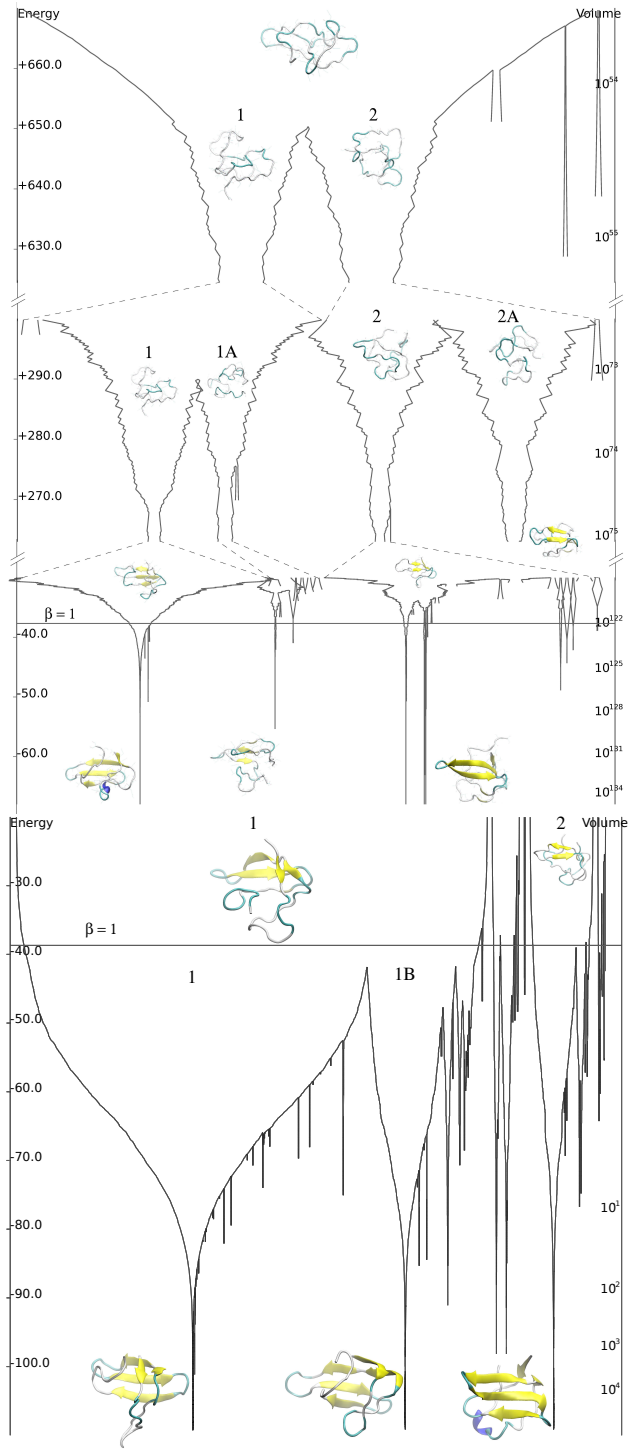

\begin{center}
\centerline{\includegraphics[width=.5\textwidth]{S4_reduced.pdf}}
\centerline{\includegraphics[width=.5\textwidth]{S4_reduced2.pdf}}
\caption{Top: prior (potential) energy landscape chart, Bottom: posterior energy landscape chart at $\beta=1$ for a simulation of the SH3 Domain using $K=15000$ and $m=15000$.   Sample conformers of funnels 1, 1A, 1B, 2 and 2A are marked on the chart.}
\label{elc4}
\end{center}
\end{figure}

The energy landscape for a simulation with $K=15000$ and $m=15000$, depicted in Fig.~\ref{elc4}, shows two main funnels (funnels 1 and 2) that further split into sub-funnels (funnels 1 and 1A, and funnels 2 and 2A).  At room temperature, conformations representative of funnels 1 and 2 are accessible, and most of the posterior mass is in funnel 1, which contains conformers with the correct backbone topology. Funnel 1 splits further at $-43$ energy units. The sub-funnels 1 and 1B are connected at room temperature, and the probability of moving from one to the other is non-zero, due to the non-zero probability of adopting a conformation that is indistinguishable from those typical to funnels 1 and 1B.  The estimated energy at room temperature is $-39$ energy units.

Fig.~\ref{fourone} shows low energy conformers from the same simulation.  Conformers 1 and 1B (from funnels 1 and 1B in Fig.~\ref{elc4}) have the correct backbone topology and a backbone RMSD of 4.61\,\AA\ and 5.47\,\AA\ with respect to the crystal structure.  Moreover, although the N-terminal loop is not included in the `regularised' contact map used in the simulation, the packing of the loop is in reasonable agreement with the crystal structure.  Conformer 2, taken from the bottom of the other major funnel, adopts a conformation with an incorrect backbone topology -- note the relative positions of the N and C termini with respect to the sheet.  The conformers shown in Fig.~\ref{fourone} correspond to lower energies than the estimated energy at room temperature, but have been shown as it is clearer to see the differences between them once the $\beta$-strands have fully formed.  Conformations available at room temperature typically have shorter $\beta$-strands.  The estimated backbone RMSD at room temperature is $\mathbb{E} ( \mathrm{RMSD} | \beta=1 ) = 6.46\,\text{\AA}$.



\subsection*{Chymotrypsin Inhibitor 2.} 
Chymotrypsin inhibitor 2 is a 65-residue protein which contains a four-stranded $\beta$-sheet and an $\alpha$-helix, but differs in topology from protein G. However, it also possesses a similar symmetry with regard to the packing of the $\alpha$-helix against the $\beta$-sheet.

As with protein G, the energy landscape shows two main folding funnels.  Fig.~\ref{elc} shows the prior energy landscape chart and the posterior energy landscape chart at room temperature for a simulation with $K=15000$ and $m=15000$.  Sample conformers from the main funnels at different energy levels are also included.  The posterior energy landscape chart shows that virtually all the posterior mass is in funnel 1 (including funnel 1A) at room temperature, with funnel 2 being insignificant.  Funnel 1 splits at $-75$ energy units into funnels 1 and 1A, which are connected at room temperature; the probability of adopting a conformation that is indistinguishable from the ones in funnels 1 and 1A is non-zero.  The estimated value of the energy at $\beta=1$ is $-82$ units.

\begin{figure}[!tpb]
\begin{center}
\centerline{\includegraphics[width=.7\textwidth]{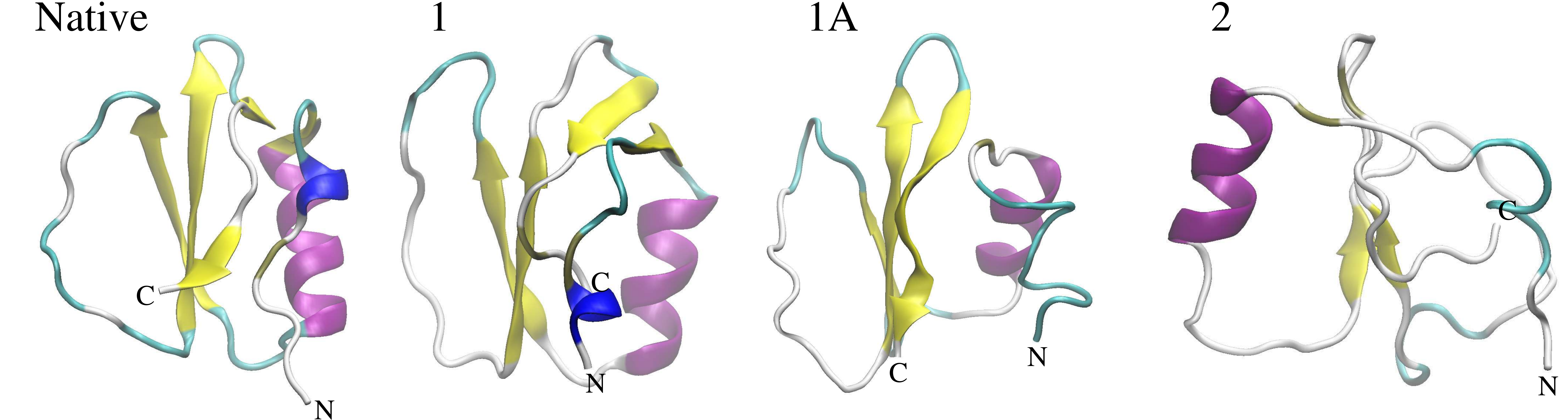}}
\caption{The native (crystal) structure for chymotrypsin inhibitor 2 (left) and 3 conformers from a simulation with $K=15000$ and $m=15000$. 1 and 1A: conformers from near the bottom of the funnel which contains conformers with the correct topology with backbone RMSD of 4.86\,\AA\ and 4.91\,\AA, respectively. The 1A conformer has the hydrophobic residues of the C-terminal $\beta$-strand on the wrong side.  2: a conformer from near the bottom of the other funnel with backbone RMSD of 11.18\,\AA.  Note that the helix is packed on the wrong side of the sheet in conformer 2.  The N and C termini are shown.}
\label{fourtwo}
\end{center}
\end{figure}

\begin{figure}[!tpb]
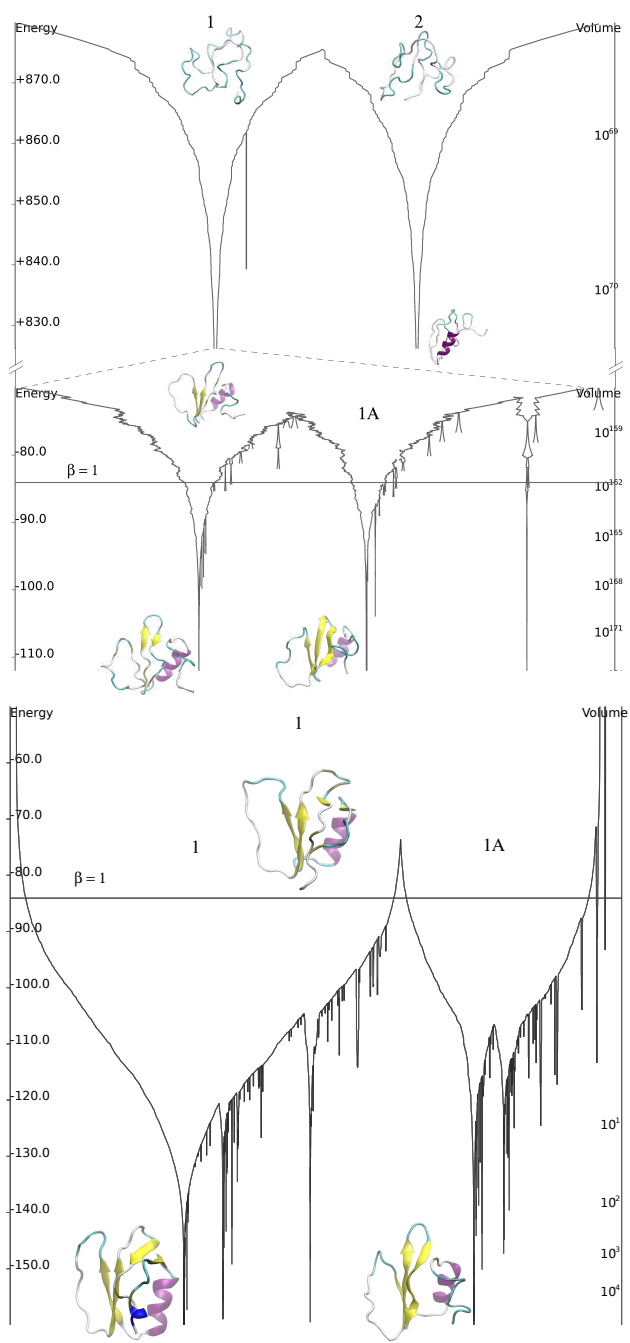

\begin{center}
\centerline{\includegraphics[width=.5\textwidth]{S6_reduced.pdf}}
\centerline{\includegraphics[width=.5\textwidth]{S6_reduced2.pdf}}
\caption{Top: prior (potential) energy landscape chart, Bottom: posterior energy landscape chart at $\beta=1$ for a simulation of the chymotrypsin inhibitor 2 using $K=15000$ and $m=15000$.  Sample conformers of funnels 1, 1A and 2 are marked on the charts.}
\label{elc}
\end{center}
\end{figure}

The native structure and sample conformers from the main funnels are also compared in Fig.~\ref{fourtwo}.
As with protein G, in our model, the helix can be packed on either side of the sheet, and in Fig.~\ref{fourtwo} conformers 1 and 2 are taken from the bottom of the two funnels. Conformer 1 has the correct backbone topology, whereas conformer 2 has the helix packed on the incorrect side of the sheet.
Conformer 1A from funnel 1A has the correct backbone topology, but the hydrophobic residues of the C-terminal $\beta$-strand are packed on the wrong side of the sheet.  Both conformers 1 and 1A have much lower energy than found at room temperature, and it is interesting to note that in both of these conformers the C-terminal $\beta$-strand has formed spontaneously without contact bias.  In the crystal structure this strand is actually a large coil. In simulations, we find that the secondary structure that is defined by  the regularised contact map  forms first, and, since the model allows a large amount of freedom for residues which do not have contact bias, nested sampling then tries to place the remaining residues in the lowest energy position possible. 
The backbone RMSD of conformers 1, 1A and 2, from the crystal structure, are 4.86\,\AA, 4.91\,\AA\ and 11.18\,\AA, respectively, while the estimated backbone RMSD at room temperature is $\mathbb{E} ( \mathrm{RMSD} | \beta=1 ) = 5.55\,\text{\AA}$.

\newpage
\section{Rigidity and Flexibility of Protein G}
\begin{figure}[!hpb]
\begin{center}
\includegraphics[width=0.5\textwidth]{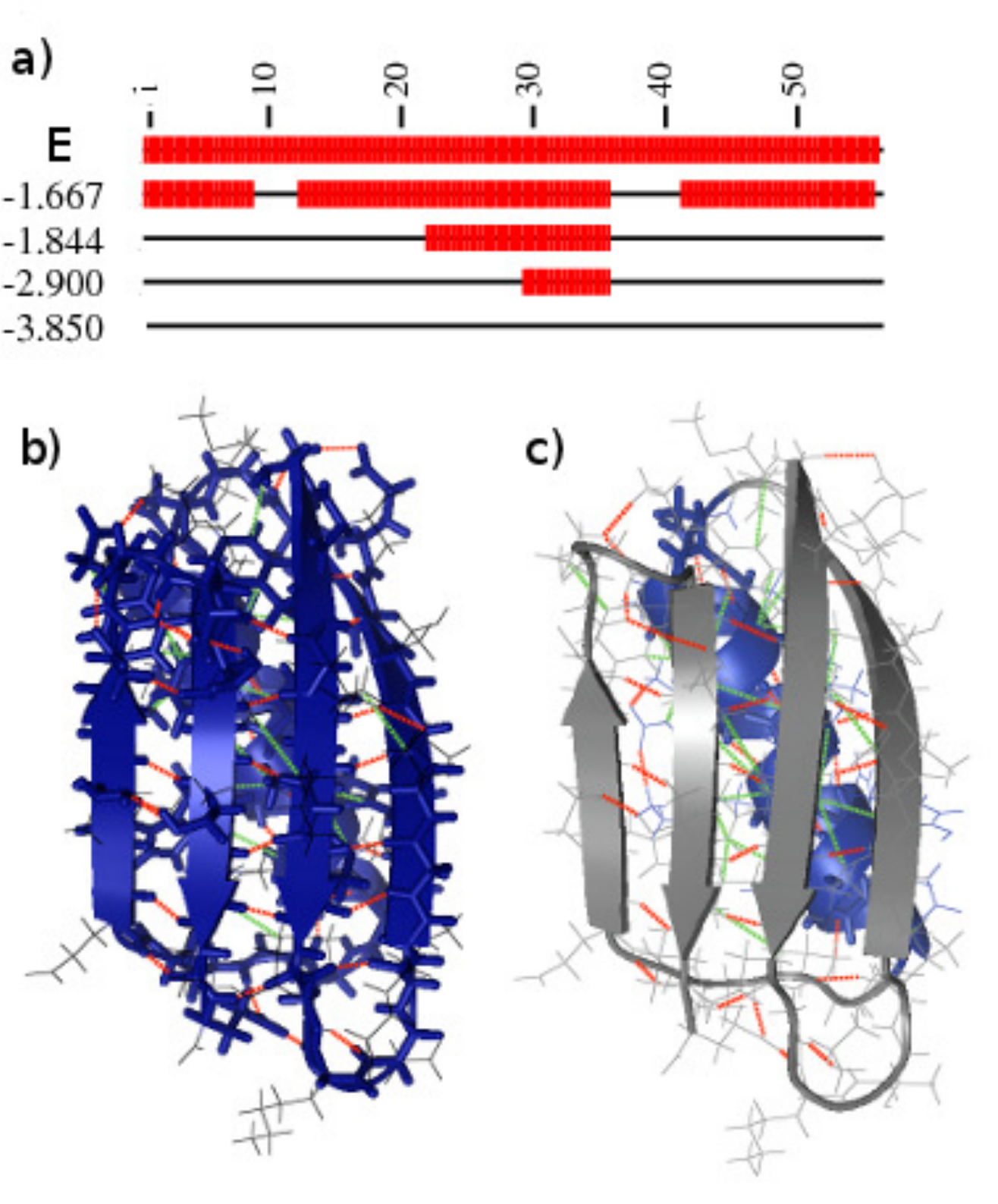}
\caption{ (a) Rigidity dilution of Protein G structure 1PGA as hydrogen bond energy cutoff is lowered. Each line represents the protein backbone; a thin line represents a flexible region while a thick line indicates membership of a large rigid cluster. The beta-sheets lose their rigidity at a cutoff of $-1.844$ kcal/mol while the helix (residues 22-35) remains rigid to much lower cutoffs. (b) Rigid cluster decomposition of 1PGA at a cutoff of $-1.0$ kcal/mol. (c) Rigid cluster decomposition at a cutoff of $-1.9$ kcal/mol. Green and red dashed lines represent hydrophobic tethers and hydrogen bonds. }
\label{fig:rcd}
\end{center}
\end{figure}

We have carried out rigidity analysis on  Protein G using the software ``FIRST'' \cite{ThoLRJK01}.    Our input was the \mbox{1PGA.pdb} structure with hydrogens added using the ``Reduce'' software \cite{WorLRR99}. Rigidity analysis balances the degrees of freedom of the atoms against the constraints introduced by covalent bonding, hydrophobic tethers, salt bridges and hydrogen bonds. The result is a decomposition of the structure into rigid and flexible regions, known as a rigid cluster decomposition. The set of hydrogen bonds included is determined by a (negative) cutoff energy $E_{cut}$. An $E_{cut}$ value near zero will include a large number of weak bonds and largely rigidify a structure; progressively lowering $E_{cut}$ eliminates hydrogen bonds in an order from weakest to strongest, a process known as ``rigidity dilution''. The progress of this loss of rigidity can be mapped in a dilution plot (Figure \ref{fig:rcd}a) in which the rigid cluster membership of each residue is mapped onto a 1-D representation of the protein backbone. A new line is plotted for each cutoff energy at which the rigidity of the mainchain changes.

At cutoff energies above $-1.844$ kcal/mol, the helix and the beta-sheet form a single rigid cluster (Figure \ref{fig:rcd}b), while at lower energies the helix is a rigid body but the beta-sheet has become flexible (Fig.~\ref{fig:rcd}c). We stress that the backbone--backbone hydrogen bonding in both the helix and the sheet persists to much lower cutoff values. Once the helix and the sheet are not a single rigid cluster, motion of the helix with respect to the sheet becomes possible. The amplitude of such motion will be constrained by covalent and non-covalent interactions, in particular the many hydrophobic tethers between helix and sheet residues.

We obtain an eigenvector for flexible motion using a coarse-grained (one site per residue) elastic network model as implemented in the software ``ElNeMo'' \cite{SuhS04}. The lowest-frequency non-trivial mode, mode 7, corresponds to a rotation of the helix about an axis perpendicular to the beta-sheet. Linear projection of the structure along this mode would rapidly introduce unphysical distortions such as elongation of the helix. In order to project the motion to finite non-zero amplitude, we make use of geometric simulation using the ``FRODA'' module \cite{WelMHT05} included in FIRST. FRODA generates new conformations of the protein structure by repeatedly introducing small perturbations of the atomic positions and reimposing the constraints. We use the elastic-network mode eigenvector to bias the perturbations \cite{JimFRW11}; this allows us to project the motion to large amplitudes while maintaining covalent, non-covalent and steric constraints.

\begin{figure}[!hpb]
\begin{center}
\includegraphics[width=0.5\textwidth]{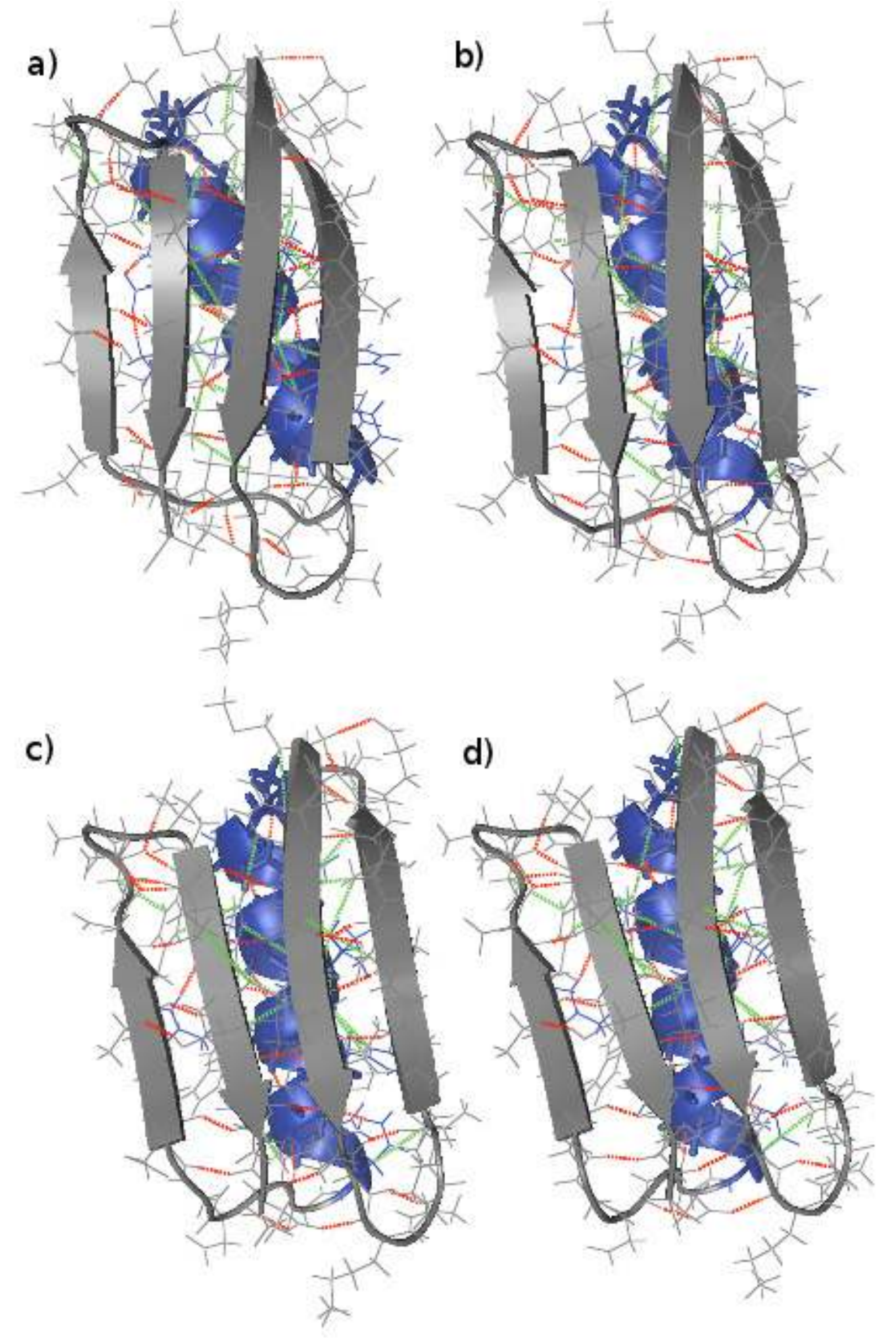}
\caption{ Projection of lowest-frequency non-trivial elastic network mode from initial structure (a) to $3\text{\AA}$ RMSD (b--d). Green and red dashed lines represent hydrophobic tethers and hydrogen bonds. }
\label{fig:flex}
\end{center}
\end{figure}

The mode can be projected to a C$_\alpha$ RMSD of more than 3 \AA\: from the initial structure (Figure \ref{fig:flex}a) while maintaining the network of hydrophobic tethers that are present in the original crystal structure. During this projection (Fig.~\ref{fig:flex}b--d) the helix rotates from its initial position diagonal to the sheet to lie parallel to the beta-sheet strands. The projected structures are very similar to conformations from the folding simulation, 
shown in the main article.   

\end{document}